\documentclass[12pt]{article}

\usepackage{graphicx}
\usepackage{amsmath}
\usepackage{cite}
\usepackage[normalem]{ulem}
\usepackage{color}
\usepackage[dvipsnames]{xcolor}
\usepackage{mathrsfs}
\usepackage{booktabs}
\usepackage{adjustbox}
\usepackage{mathtools}
\usepackage{enumerate}
\usepackage{slashed}

\usepackage{amssymb,esint}
\usepackage{amsthm}
\usepackage{enumitem}
\usepackage{float}
\usepackage[utf8]{inputenc}
\usepackage[T1]{fontenc}
\usepackage[margin=1in]{geometry}

\def\beq{\begin{equation}}
\def\eeq{\end{equation}}
\def\bea{\begin{eqnarray}}
\def\eea{\end{eqnarray}}
\def\nn{\nonumber}

\def\roughly#1{\mathrel{\raise.3ex\hbox
{$#1$\kern-.75em\lower1ex\hbox{$\sim$}}}}

\def\sla#1{\raise.15ex\hbox{$/$}\kern-.57em #1}% Feynman slash

\newcommand{\op}[3]{\O^{#2,#3}_{#1}}
\renewcommand{\O}{\mathcal{O}}
\newcommand{\hc}{\mathrm{h.c.}}

\begin{document}

\begin{flushright}
UdeM-GPP-TH-22-295 \\
\end{flushright}

\begin{center}
\bigskip
{\Large \bf \boldmath Dimension-8 SMEFT Matching Conditions for the
  \\[0.3cm] Low-Energy Effective Field Theory} \\
\bigskip
\bigskip
{\large
Serge Hamoudou $^{a,}$\footnote{Present address: Physics and Astronomy, McMaster University, 1280 Main St W, Hamilton,
ON L8S 4M6, Canada. E-mail: hamoudos@mcmaster.ca.},
Jacky Kumar $^{a,b,}$\footnote{Present address: Theoretical Division, MS B283, Los Alamos National Laboratory, Los Alamos,
NM 87545, U.S.A. E-mail: jacky.kumar@lanl.gov} \\
and David London $^{a,}$\footnote{london@lps.umontreal.ca}}
\end{center}

\begin{flushleft}
~~~~~~~~~~~~~~~~$a$: {\it Physique des Particules, Universit\'e de Montr\'eal,}\\
~~~~~~~~~~~~~~~~~~~~~{\it 1375 Avenue Th\'er\`ese-Lavoie-Roux, Montr\'eal, QC, Canada  H2V 0B3} \\
~~~~~~~~~~~~~~~~$b$: {\it Institute for Advanced Study, Technical University Munich, }\\
~~~~~~~~~~~~~~~~~~~~{\it  Lichtenbergstr.\ 2a, D-85747 Garching, Germany}
\end{flushleft}

\begin{center}
\bigskip (\today)
\vskip0.5cm {\Large Abstract\\} 
\vskip3truemm
\parbox[t]{\textwidth}{In particle physics, the modern view is to categorize things in terms of effective field theories (EFTs). Above the weak scale, we have the SMEFT, formed when the heavy new physics (NP) is integrated out, and for which the Standard Model (SM) is the leading part. Below $M_W$, we have the LEFT (low-energy EFT), formed when the heavy SM particles ($W^\pm$, $Z^0$, $H$, $t$) are also integrated out. In order to determine how low-energy measurements depend on the underlying NP, it is necessary to compute the matching conditions of LEFT operators to SMEFT operators. These matching conditions have been worked out for all LEFT operators up to dimension 6 in terms of SMEFT operators up to dimension 6 at the one-loop level. However, this is not sufficient for all low-energy observables. In this paper we present the {momentum-independent} matching conditions of all such LEFT operators to SMEFT operators up to dimension 8 {at tree level}.}
\end{center}

\thispagestyle{empty}
\newpage
\setcounter{page}{1}
% Decrease texheight (for preprint numbers) again
%\textheight 23.0 true cm
\baselineskip=14pt
\tableofcontents

\newpage 

\section{Introduction}

Despite its enormous success in accounting for almost all experimental
data to date, the Standard Model (SM) of particle physics still has no
explanation for a number of other key observations, such as neutrino
masses, the baryon asymmetry of the universe, dark matter, etc. For
this reason, it is widely believed that there must exist physics
beyond the SM. And since the LHC has not discovered any new particles
up to a scale of $O({\rm TeV})$, this new physics (NP) is likely to be
very massive.

When the NP is integrated out, one obtains an effective field theory
(EFT), of which it is now generally believed that the SM is simply the
leading part. This EFT must obey the SM gauge symmetry $SU(3)_C \times
SU(2)_L \times U(1)_Y$. Since the discovery of the Higgs boson, the
default assumption is that this symmetry is realized linearly, i.e.,
the symmetry is broken via the Higgs mechanism, resulting in the
Standard Model EFT, or SMEFT (see, {\it e.g.},
Refs.~\cite{Buchmuller:1985jz, Brivio:2017vri}). The SMEFT has been
studied extensively: a complete and non-redundant list of dimension-6
operators is given in Ref.~\cite{Grzadkowski:2010es}, the dimension-7 operators can be found in Ref.~\cite{Lehman, XiaoDongMa}, and the dimension-8 operators are tabulated in Refs.~\cite{Li:2020gnx,
  Murphy:2020rsh}.

The LEFT (low-energy effective field theory) describes the physics
below the $W$ mass, and is produced when the heavy SM particles ($W$,
$Z$, $t$, $H$) are also integrated out. (This is also called the WET
(weak effective field theory).) In Ref.~\cite{Jenkins:2017jig},
Jenkins, Manohar and Stoffer (JMS) present a complete and
non-redundant basis of LEFT operators up to dimension 6, including
those that violate $B$ and $L$, and also give the matching to dimension-6 SMEFT
operators at tree level. 

{At one loop, there are two additional types of contributions to the matching. First, there are the renormalization-group running effects which are enhanced by $\log{(\mu_{\rm NP} / \mu_{\rm EW})}$. Second, we have the threshold corrections, or one-loop matching at the electroweak scale, which are a part of the next-to-leading-log effects. Naively, the former contributions appear to dominate. However, for certain processes, the latter can be comparable \cite{Bobeth:2017xry}.
The anomalous-dimension matrices for the renormalization-group running  and the one-loop matching effects for the full SMEFT and LEFT bases have been computed in Refs.~\cite{Alonso:2013hga} and \cite{Dekens:2019ept}, respectively. (Note that, in order it to consistently include the threshold corrections at the electroweak scale, it is essential to have two-loop anomalous dimensions of the LEFT operators \cite{Aebischer:2021raf, Aebischer:2022tvz}.)}

With this information, if a discrepancy with the SM
is observed in a process that uses a particular LEFT operator, we will
know which dimension-6 SMEFT operators are involved.

However, this is not always sufficient. Information about the
contributions from higher-dimension operators may be important if the
process in question is suppressed in the SM and/or is very precisely
measured. Examples of observables for which such contributions must be taken into account include electroweak precision data from
LEP \cite{Corbett:2021eux}, lepton-flavour-violating processes
\cite{Ardu:2021koz, Ardu:2022pzk}, meson-antimeson mixing ($\Delta F = 2$)
\cite{Bobeth:2017xry, Silva:2022tln}, and electric dipole moments
\cite{Panico:2018hal}. (Dimension-8 SMEFT operators have also been
discussed in the context of high-energy processes, see
Refs.~\cite{Chala:2018ari, Hays:2018zze, Ellis:2019zex,
  Alioli:2020kez, Hays:2020scx, Ellis:2020ljj, Boughezal:2021tih,
  Gu:2020ldn}.) 

{We have the matching conditions of dimension-6 LEFT operators to dimension-6 SMEFT operators. A first step is therefore to extend this matching to include the (subdominant) dimension-8 SMEFT operators. But there is a complication: dimension-8 SMEFT operators will also produce dimension-8 LEFT operators. (A complete set of dimension-8 LEFT operators can be found in Ref.~\cite{Murphy:2020cly}.) Thus, additional LEFT operators must in principle also be considered.}

{Note that a distinction can be made between the various contributions. Consider four-fermion operators. These dimension-6 LEFT operators have no derivatives, and are therefore momentum-independent (MI). On the other hand, the dimension-8 extensions of four-fermion operators do contain derivatives, i.e., they are momentum-dependent (MD). As a consequence, their tree-level matching conditions to dimension-8 SMEFT operators are also MD\footnote{In a similar vein, one can find MD contributions to dimension-7 LEFT operators due to dimension-6 and 7 SMEFT operators, see Ref.~\cite{Liao:2020zyx} .}. Of course, the MD contributions can be at the same level in power counting as the MI contributions. Depending on the scale of the NP ($\Lambda$) and the masses of the fermions involved in the process under consideration, they can be numerically comparable to, or even larger than, the MI contributions. Thus, a full computation of the contributions from higher-dimension operators to low-energy processes must include both dimension-6 and dimension-8 LEFT operators and their MI and MD matching conditions to dimension-8 SMEFT operators. This is an enormous undertaking.}

{Fortunately, the MI and MD tree-level matching conditions can be separated. In the present paper, we focus only on the MI matching conditions. MD matching conditions will be presented elsewhere. 
Note that a complete analysis of the relationship between LEFT operators and dimension-8 SMEFT operators must also take into account the renormalization-group running of SMEFT operators from the NP scale down to low energy, as well as the threshold corrections at the electroweak scale. For bosonic SMEFT operators up to dimension 8, the anomalous dimensions have been calculated in Refs.~\cite{MikaelChala1, MikaelChala2}.}

In our analysis, we follow closely the approach of
Ref.~\cite{Jenkins:2017jig}, and extend it to include dimension-8
SMEFT operators.  Below, we often refer to this paper simply by the
initials of its authors, as JMS.

We begin in Sec.~2 with some preliminary remarks comparing our
analysis with that of JMS, and discuss in general terms how matching
conditions are computed. In Sec.~3, we present the setup, showing how
the presence of dimension-8 SMEFT operators affects the symmetry
breaking, the generation of masses, and the couplings of the gauge and
Higgs bosons to fermions. The computations required to derive the complete matching conditions are described in Sec.~4. {Although we do not compute the MD matching conditions, the various sources of such contributions are outlined here.} We conclude in Sec.~5. The results are presented in Appendix D. 
Appendices A, B, C 
give a variety of information relevant to the details of the analysis.

\section{Preliminaries}
\label{Preliminaries}

In Ref.~\cite{Jenkins:2017jig}, JMS compute the tree-level SMEFT matching conditions for the LEFT operators. The matching conditions for operators that conserve both $B$ and $L$ involve only even-dimension SMEFT operators, and are given up to dimension 6. For operators that violate $B$ and/or $L$, the matching conditions can involve even- or odd-dimension SMEFT operators (but not both), depending on the operator, and are computed to dimension 6 or dimension 5. In the present paper, we extend this analysis: we compute these matching conditions up to dimension 8 (dimension 7) if even-dimension (odd-dimension) SMEFT operators are involved. (In this paper, when we refer to ``computing the matching conditions up to dimension 8,'' both of these possibilities are understood.) If one eliminates the dimension-8 or dimension-7 contributions, the results of JMS are reproduced. This makes it easy to compare the results. Also, we present the elements of our analysis in much the same order as JMS.

In the LEFT Lagrangian, we consider only operators up to dimension 6 (like JMS):
\beq
\mathcal L_{\text{LEFT}}=\mathcal L_{\text{LEFT}}^{\text{Neutrino mass}}+\mathcal L_{\text{QCD}+\text{QED}}+\sum_{n=5}^6 \, \sum_{\mathcal O\in\dim n}\dfrac{C_{\mathcal O}} {\Lambda^{n-4}}\,\mathcal O ~.
\eeq
For the SMEFT, all operators up to dimension 8 are included:
\beq
\mathcal L_{\text{SMEFT}}=\mathcal L_{\text{SM}}+ \sum_{n=5}^8 \, \sum_{Q\in\dim n}\dfrac{C_Q}{\Lambda^{n-4}}\,Q ~.
\eeq
{(Note that using the same suppression scale $\Lambda$ for both LEFT and SMEFT is just a matter of convention.)}

Still, there are two differences in our notation:
\begin{itemize}

\item Our convention is to have dimensionless Wilson coefficients
  (WCs). For instance, for the dimension-6 SMEFT lagrangian, we write
\beq
\mathcal L^{(6)}_{\text{SMEFT}}=\sum_{Q\in\dim6}\dfrac{C_{Q}}
{\Lambda^2}\,Q ~.
\eeq
This convention is different from that of JMS, which uses dimensionful
WCs.

\item In the unbroken phase, the SM lagrangian is
\bea
\mathcal L_{\text{SM}} &=& \, -\dfrac14 G^A_{\mu\nu} G^{A\mu\nu} - \dfrac14 W^I_{\mu\nu} W^{I\mu\nu}
- \dfrac14 B_{\mu\nu} B^{\mu\nu} \nn\\
&& +~\sum_{\psi=q,u,d,l,e} \overline\psi i\slashed D\psi + (D_\mu H)^\dagger (D^\mu H)
- \lambda \left(H^\dagger H-\dfrac12v^2\right)^2 \nn\\
&& -~\left[\overline l_pe_r(Y_e)_{pr}H+\overline q_pu_r(Y_u)_{pr}\tilde H+\overline q_pd_r(Y_d)_{pr}H+\text{h.c.}\right]
\nn\\
&& +~\dfrac{\theta_3{g_s^2}}{32\pi^2}G^A_{\mu\nu}\tilde G^{A\mu\nu}
+\dfrac{\theta_2g^2}{32\pi^2}W^I_{\mu\nu}\tilde W^{I\mu\nu}+\dfrac{\theta_1{g'}^2}{32\pi^2}B_{\mu\nu}\tilde B^{\mu\nu} ~.
\label{SMLagrangian}
\eea
This uses the same notation as JMS, with one exception: our Yukawa
matrices (the $Y$s) are the hermitian conjugates of those of JMS.

\end{itemize}

In Eq.~(\ref{SMLagrangian}), the fields $q_r$ and $l_r$ are
(left-handed) $SU(2)_L$ doublets, while $u_r$, $d_r$ and $e_r$ are
(right-handed) $SU(2)_L$ singlets, where $r = 1,~2,~3$ is a generation
(weak-eigenstate) index. The physical (mass-eigenstate) states are the
same for the charged leptons, the left- and right-handed $u$-type
quarks, and the right-handed $d$-type quarks. For the left-handed
$d$-type quarks, the relation between the weak and mass eigenstates is
\beq
d_{Lr} = V_{rd} d_L + V_{rs} s_L + V_{rb} b_L \equiv V_{rx} d_{Lx} ~,
\eeq
where the left-hand side is a weak eigenstate, and the right-hand side
is a linear combination of mass eigenstates. The $V_{rx}$ are elements
of the unitary mixing matrix, which is the Cabibbo-Kobayashi-Maskawa
(CKM) matrix in the SM. Note: as in JMS, our LEFT matching conditions are given in the weak eigenstate basis. They can be written in terms of the physical states by using the above relation.

In our analysis, we make reference to several different sets of
operators. The LEFT operators are taken from JMS
\cite{Jenkins:2017jig}, the dimension-6 SMEFT operators are found in
Ref.~\cite{Grzadkowski:2010es}, and we use Ref.~\cite{Murphy:2020rsh}
for the dimension-8 SMEFT operators. In all cases, we use the same
notation for the operators and their WCs as is used in the references.
For the dimension-7 SMEFT operators, we use a basis that is equivalent that of Ref.~\cite{Lehman}, but with a different notation. For convenience, in the Appendices, we present tables of all the operators
used in this paper. These include LEFT operators (Appendix A), along with
dimension-5 to 8 SMEFT operators (Appendix B).

It is useful to give an example that illustrates the various issues
involved in deriving matching conditions. Consider the charged-current
four-fermion operator
\beq
{\mathcal O}_{\nu edu}^{V,LL} = ({\bar\nu}_{Lp}\gamma^\mu e_{Lr})({\bar d}_{Ls}\gamma_\mu u_{Lt}) + {\rm h.c.},
~~~~~ {\rm coefficient:} \dfrac1{\Lambda^2}C_{\substack{\nu edu\\prst}}^{V,LL} ~.
\eeq
We begin by examining the matching to the SM. That is, ${\mathcal
  O}_{\nu edu}^{V,LL}$ is taken to be an operator of the Fermi theory,
whose coefficient has magnitude $4 G_F / \sqrt{2}$. The SM Lagrangian
consists only of operators of at most dimension 4. This four-fermion
operator can be generated in the SM when a $W$ is exchanged between
the two fermion currents, and the $W$ is integrated out. The SM
matching condition is then
\beq
\dfrac1{\Lambda^2}C_{\substack{\nu edu\\prst}}^{V,LL} = -\dfrac{g^2}{2M_W^2}\,[W_l]_{pr}[W_q]_{ts}^* ~.
\label{SMmatch}
\eeq
Here, $W_l$ and $W_q$ are the couplings of the $W$ to the lepton and
quark pair, respectively. In the {weak-eigenstate basis of the} SM, $[W_l]_{pr} = \delta_{pr}$ and
$[W_q]_{ts} = \delta_{ts}$. Knowing that the coefficient has magnitude
$4 G_F / \sqrt{2}$, this leads to the well-known relation
\beq
\frac{G_F}{\sqrt{2}} = \dfrac{g^2}{8M_W^2} ~.
\eeq

The matching to SMEFT at dimension 6 was computed by JMS. It is
\beq
\dfrac1{\Lambda^2} C_{\substack{\nu edu\\prst}}^{V,LL} \, + {\rm h.c.} = \dfrac2{\Lambda^2} C^{(3)}_{\substack{lq\\prst}}
- \dfrac{\overline g^2}{2M_W^2} \, [W_l]^{\text{eff}}_{pr}{[W_q]^{\text{eff}}_{ts}}^* + \, {\rm c.c.}
\label{SMEFTdim6match}
\eeq
Since the SMEFT includes dimension-6 terms, it contains the
four-fermion operator. That is, there is a {\it direct} contribution
to the matching conditions, $C^{(3)}_{\substack{lq\\prst}}$. As was
the case in the SM, $C_{\substack{\nu edu\\prst}}^{V,LL}$ can also be
generated by the exchange of a $W$ between the two fermion
currents. This is represented by the second term above. Although this
resembles the term in Eq.~(\ref{SMmatch}), there are several
differences:
\begin{enumerate}

\item In the presence of dimension-6 SMEFT operators, the coupling
  constant is modified: $g \to {\bar g}$. This is due to the fact that, when one adds dimension-6 corrections to the
  kinetic terms of the gauge bosons, these fields and the coupling constants must be redefined in
  order to ensure that the kinetic term is properly normalized.

\item In the SM, the $W$ coupling to fermions is fixed by the fermion
  kinetic term, ${\bar\psi} {\slashed D} \psi$. In SMEFT, there are
  dimension-6 corrections, such as {$H^\dagger i D_\mu H {\bar\psi} \gamma^\mu \psi$}. These will change the magnitudes of the couplings, and
  permit inter-generational couplings, hence the `eff' superscript on
  $W_l$ and $W_q$.

\end{enumerate}
The bottom line is that many dimension-6 SMEFT operators are
implicitly present in the second term of Eq.~(\ref{SMEFTdim6match})
above. Collectively, these operators form the {\it indirect}
contributions. They must be carefully taken into account in the
matching conditions. (Note that, if one expands the effective parameters appearing in the matching conditions, many terms will appear; those that are of higher order than dimension 8 are to be ignored.)

\section{Setup}
\label{Setup}

The Lagrangian for the SM in the unbroken phase is given in
Eq.~(\ref{SMLagrangian}). When the Higgs field acquires a vacuum expectation value (vev), given
by the minimum of the Higgs potential, the symmetry is broken, and
masses are generated for the $W^\pm$, the $Z^0$ and the fermions. One
can easily compute the masses of the physical gauge bosons, as well as
their couplings to the physical fermions, in terms of the parameters
of $\mathcal L_{\text{SM}}$, in particular $g$, $g'$ and $v$.

When one includes higher-order SMEFT operators of dimension 6, 8,
etc., this whole process must be recalculated in order to take into
account these new operators. One must make field redefinitions so that
the kinetic terms are properly normalized, the minimum of the Higgs
potential (i.e., the Higgs vev) must be recomputed, corrections to
$\sin\theta_W$ must be taken into account, etc. One sees the effects
of these additional operators in the redefinitions of the coupling
constants, the couplings of gauge bosons to fermions, and other
quantities that appear in both the direct and indirect contributions
to the matching conditions.

In this section, we present the main effects of including SMEFT
operators up to dimension 8. We emphasize those results that are
important for the matching conditions. These results are in agreement with the predictions of the geometric formulation of the SMEFT \cite{geomSMEFT}.

\subsection{Higgs sector}

After the Higgs acquires a vev, we redefine the Higgs field as
follows:
\beq
H\to\dfrac1{\sqrt2}\begin{pmatrix} 0 \\ [1+c_{H,\text{kin}}]h+v_T \end{pmatrix} ~.
\eeq
Here, $v_T$ and $c_{H,\text{kin}}$ are respectively determined by
minimizing the Higgs potential and by normalizing the Higgs kinetic
term.

\subsubsection{Higgs vev}

In the presence of SMEFT operators up to dimension 8, the Higgs
potential is
\beq
V(H) = \lambda\left(H^2-\dfrac12v^2\right)^2-\dfrac1{\Lambda^2}C_{H}H^6-\dfrac1{\Lambda^4}C_{H^8}H^8 ~,
\eeq
where only the real part of the second component of the Higgs doublet,
$H$, is taken to be nonzero. We define the physical Higgs vev, $v_T$,
to be $v_T \equiv \sqrt2\,H^{\text{min}}$, where $H^{\text{min}}$
minimizes the Higgs potential.  This implies that
\beq
\label{Higgs' vev dimension 8}
v_T = v \left(1+\dfrac{3v^2}{8\lambda\Lambda^2}\,C_{H}
+\dfrac{v^4}{4\lambda\Lambda^4}\left[\dfrac{63}{32\lambda}\,[C_H]^2+C_{H^8}\right]\right) ~.
\eeq
$v_T$ is the physical parameter that appears in the matching
relations, and whose value can in principle be determined by a fit to
the data.

\subsubsection{Higgs kinetic term}

Including SMEFT contributions up to dimension 8, the Higgs kinetic
term is
\beq
\mathcal L_{\text{SMEFT}}^{\text{Higgs kinetic}} = \,\,\dfrac12\left[1+\dfrac{2 v_T^2}{\Lambda^2}
  \left(\dfrac14C_{HD}-C_{H\Box}\right)+\dfrac{v_T^4}{4\Lambda^4}
  \left(C_{H^6}^{(1)}+C_{H^6}^{(2)}\right)\right](1+c_{H,\text{kin}})^2(\partial_\mu h)(\partial^\mu h) ~.
\eeq
In order for this term to be properly normalized, one must have
\beq
\label{Higgs scalar field term normalization}
c_{H,\text{kin}}=\dfrac{v_T^2}{\Lambda^2}\left(C_{H\Box}-\dfrac14C_{HD}\right)-\dfrac{v_T^4}{8\Lambda^4}\left(C_{H^6}^{(1)}+C_{H^6}^{(2)}\right)+\dfrac{3 v_T^4}{2\Lambda^4}\left(C_{H\Box}-\dfrac14C_{HD}\right)^2 ~.
\eeq
This is essentially a redefinition of the normalization of the Higgs
field.

\subsubsection{Higgs mass}

Taking into account the SMEFT contributions up to dimension 8, the Higgs boson mass term is
\beq
\mathcal L_{\text{SMEFT}}^{\text{Higgs mass}}=\dfrac12\left[\lambda v^2-3\lambda{v_T^2}+\dfrac{15{v_T^4}}{4\Lambda^2}\,C_{H}+\dfrac{7{v_T^6}}{2\Lambda^4}\,C_{H^8}\right](1+c_{H,\text{kin}})^2h^2 ~.
\eeq
This gives the following expression for the Higgs boson mass:
\beq
\label{Higgs boson masses dimension 8}
{m_h}^2=(1+c_{H,\text{kin}})^2{v_T^2}\left[2\lambda-\dfrac{3{v_T^2}}{\Lambda^2}\,C_{H}-\dfrac{3{v_T^4}}{\Lambda^4}\,C_{H^8}\right] ~.
\eeq

\subsection{Fermion mass matrices \& Yukawa couplings}

Before symmetry breaking, the SMEFT Lagrangian up to dimension 8
contains the following terms for charged leptons and quarks:
\beq
-(Y_\psi)_{pr} \, {\bar\chi}_{p} \psi_{r} \, \overline H
+ \frac{1}{\Lambda^2} C_{\substack{\psi H\\pr}} \, {\bar\chi}_{p} \psi_{r} \, \overline H (H^\dagger H)
+ \frac{1}{\Lambda^4} C_{\substack{\chi\psi H^5\\pr}} \, {\bar\chi}_{p} \psi_{r} \, \overline H (H^\dagger H)^2 + {\rm h.c.} ~,
\eeq
where $\psi\in\{e,u,d\}$ (right-handed $SU(2)_L$ singlets),
$\chi\in\{l,q\}$ (left-handed $SU(2)_L$ doublets), $\overline H=H$ if $\psi\in\{e,d\}$ and $\overline H=\tilde H$ if $\psi\in\{u\}$. Here, the first
term (dimension 4) belongs to the SM and the last two terms are SMEFT
operators (respectively dimension 6 and 8). Lepton-number-violating terms are also present:
\beq
\frac{1}{\Lambda} C_{\substack{5\\pr}} \, \epsilon^{ij}\epsilon^{kl}(l^T_{ip}Cl_{kr}) \, H_jH_l
+ \frac{1}{\Lambda^3} C_{\substack{l^2H^4\\pr}} \, \epsilon^{ij}\epsilon^{kl}(l^T_{ip}Cl_{kr}) \, H_jH_l (H^\dagger H) + {\rm h.c.} ~.
\eeq
When the Higgs gets a vev,
both mass matrices and Yukawa coupling terms are generated.

\subsubsection{Fermion mass matrices}

The SMEFT mass terms for charged leptons and quarks up to dimension 8 are
\bea
\mathcal L_{\text{SM}}^{\text{Fermion mass}} &=& -\dfrac{v_T}{\sqrt2} \left[
(Y_e)_{pr}\overline e_{Lp}e_{Rr} + (Y_u)_{pr}\overline u_{Lp}u_{Rr} + (Y_d)_{pr}\overline d_{Lp}d_{Rr} \right] + \hc, \nn\\
\mathcal L_{\text{SMEFT},6}^{\text{Fermion mass}} &=& \dfrac{{v_T^3}}{2\sqrt2\,\Lambda^2} \left[
  C_{\substack{eH\\pr}}\overline e_{Lp}e_{Rr} + C_{\substack{uH\\pr}}\overline u_{Lp}u_{Rr}
  + C_{\substack{dH\\pr}}\overline d_{Lp}d_{Rr} \right] + \hc, \\
\mathcal L_{\text{SMEFT},8}^{\text{Fermion mass}} &=& \dfrac{{v_T^5}}{4\sqrt2\,\Lambda^4} \left[
C_{\substack{leH^5\\pr}}\overline e_{Lp}e_{Rr} + C_{\substack{quH^5\\pr}}\overline u_{Lp}u_{Rr}
+ C_{\substack{qdH^5\\pr}}\overline d_{Lp}d_{Rr} \right] + \hc \nn
\eea
This gives the following mass matrices:
\beq
\label{mass matrices dimension 8}
[M_\psi]_{pr}=\dfrac{v_T}{\sqrt2}\left[(Y_\psi)_{pr}
- \dfrac{{v_T^2}}{2\Lambda^2}\,C_{\substack{\psi H\\pr}}-\dfrac{{v_T^4}}{4\Lambda^4}\,C_{\substack{\chi\psi H^5\\pr}}\right] ~.
\eeq
The SMEFT neutrino mass terms are
\bea
\mathcal L_{\text{SMEFT}}^{\text{Neutrino mass}} &=& \dfrac{{v_T}^2}{2\Lambda} \left[C_{\substack{5\\pr}} + \dfrac{{v_T}^2}{2\Lambda^2}C_{\substack{l^2H^4\\pr}}\right]\overline\nu^T_{Lp}C\nu_{Lr} + \hc ~.
\eea
This gives the following mass matrices:
\beq
\label{neutrino mass matrices dimension 8}
[M_\nu]_{pr}=-\dfrac{{v_T}^2}{\Lambda}\left[C_{\substack{5\\pr}}+\dfrac{{v_T}^2}{2\Lambda^2}C_{\substack{l^2H^4\\pr}}\right] ~.
\eeq

\subsubsection{Yukawa couplings}

The SM, dimension-6 and dimension-8 SMEFT Yukawa coupling terms for charged leptons and quarks are
\bea
\mathcal L_{\text{SM}}^{\text{Yukawa}} &=& -\dfrac{(1+c_{H,\text{kin}})}{\sqrt2} \left[
  (Y_e)_{pr}\overline e_{Lp}e_{Rr}h + (Y_u)_{pr}\overline u_{Lp}u_{Rr}h + (Y_d)_{pr}\overline d_{Lp}d_{Rr}h \right]
+ \hc, \nn\\
\mathcal L_{\text{SMEFT},6}^{\text{Yukawa}} &=& \dfrac{3(1+c_{H,\text{kin}}) v_T^2}{2\sqrt2\,\Lambda^2} \left[
  C_{\substack{eH\\pr}}\overline e_{Lp}e_{Rr}h + C_{\substack{uH\\pr}}\overline u_{Lp}u_{Rr}h
+ C_{\substack{dH\\pr}}\overline d_{Lp}d_{Rr}h \right] + \hc, \\
\mathcal L_{\text{SMEFT},8}^{\text{Yukawa}} &=& \dfrac{5(1+c_{H,\text{kin}}) v_T^4}{4\sqrt2\,\Lambda^4} \left[
  C_{\substack{leH^5\\pr}}\overline e_{Lp}e_{Rr}h + C_{\substack{quH^5\\pr}}\overline u_{Lp}u_{Rr}h
  + C_{\substack{qdH^5\\pr}}\overline d_{Lp}d_{Rr}h \right] + \hc \nn
\eea
This gives the following Yukawa couplings (up to dimension 8):
\bea
\label{Yukawa couplings dimension 8}
(Y_\psi)^{\text{eff}}_{pr} = \dfrac{1+c_{H,\text{kin}}}{\sqrt2}\left[\dfrac{\sqrt2}{v_T}\,[M_\psi]_{pr}-\dfrac{{v_T^2}}{\Lambda^2}\,C_{\substack{\psi H\\pr}}-\dfrac{{v_T^4}}{\Lambda^4}\,C_{\substack{\chi\psi H^5\\pr}}\right] ~.
\eea
There are also momentum-dependent Yukawa couplings (i.e., ordinary Yukawa couplings with additional derivatives {acting on the Higgs field}) occurring at dimension 6 in SMEFT. However, as discussed in the introduction, MD contributions to the matching conditions are not included in the present work (though they are briefly discussed in Sec.~\ref{MDcont}). 

The SMEFT Yukawa coupling terms for neutrinos are
\bea
\mathcal L_{\text{SMEFT}}^{\text{Neutrino Yukawa}} &=& \dfrac{v_T}{\Lambda}(1+c_{H,\text{kin}})\left[C_{\substack{5\\pr}} + \dfrac{{v_T}^2}{\Lambda^2}C_{\substack{l^2H^4\\pr}}\right]h\,\overline\nu^T_{Lp}C\nu_{Lr} + \hc ~.
\eea
This gives the following Yukawa couplings:
\bea
\label{neutrino Yukawa couplings dimension 8}
(Y_\nu)^{\text{eff}}_{pr} = (1+c_{H,\text{kin}})\left[\dfrac1{v_T}\,[M_\nu]_{pr}-\dfrac{{v_T}^3}{2\Lambda^3}C_{\substack{l^2H^4\\pr}}\right] ~.
\eea
These Yukawa couplings enter the matching conditions of certain
four-fermion operators in LEFT. 

\subsection{Electroweak gauge boson masses \& mixing and coupling constants}

\subsubsection{Kinetic terms}
\label{GBkinetic}

Including the SMEFT contributions up to dimension 8, the kinetic terms
of the electroweak gauge bosons after symmetry breaking are
\beq
\mathcal L_{\text{SMEFT}}^{\text{Electroweak kin}}=-\dfrac14\left\{\begin{aligned}
& \left[1-\dfrac{2 v_T^2}{\Lambda^2}\,C_{HW}-\dfrac{ v_T^4}{\Lambda^4}\,C_{W^2H^4}^{(1)}\right]W^I_{\mu\nu}W^{I\mu\nu} \\
& -\dfrac{ v_T^4}{\Lambda^4}\,C_{W^2H^4}^{(3)}\,W^3_{\mu\nu}W^{3\mu\nu} \\
& +\left[1-\dfrac{2 v_T^2}{\Lambda^2}\,C_{HB}-\dfrac{ v_T^4}{\Lambda^4}\,C_{B^2H^4}^{(1)}\right]B_{\mu\nu}B^{\mu\nu} \\
& +\left[\dfrac{2 v_T^2}{\Lambda^2}\,C_{HWB}+\dfrac{ v_T^4}{\Lambda^4}\,C_{WBH^4}^{(1)}\right]W^3_{\mu\nu}B^{\mu\nu}
\end{aligned}\right\} ~.
\label{gaugemassterms}
\eeq
Here there are two issues that must be resolved. First, the kinetic
terms must be properly normalized. Second, the
$W^3_{\mu\nu}B^{\mu\nu}$ mixing term must be removed.

Proper normalization of the kinetic terms can be achieved by
redefining the coupling constants and the normalization of the gauge
fields:
\bea
\label{ccredefs}
\overline g &=& \left[1+\dfrac{{v_T^2}}{\Lambda^2}\,C_{HW}
  +\dfrac{{v_T^4}}{2\Lambda^4}\,C_{W^2H^4}^{(1)}+\dfrac{3{v_T^4}}{2\Lambda^4}\,[C_{HW}]^2\right]g ~,  \nn\\
\overline g' &=& \left[1+\dfrac{{v_T^2}}{\Lambda^2}\,C_{HB}
  +\dfrac{{v_T^4}}{2\Lambda^4}\,C_{B^2H^4}^{(1)}+\dfrac{3{v_T^4}}{2\Lambda^4}\,[C_{HB}]^2\right]g' ~, \\
\label{gbredefs}
W^I_\mu &=& \left[1+\dfrac{{v_T^2}}{\Lambda^2}\,C_{HW}+\dfrac{{v_T^4}}{2\Lambda^4}\,C_{W^2H^4}^{(1)}
  +\dfrac{3{v_T^4}}{2\Lambda^4}\,[C_{HW}]^2\right]\mathcal W^I_\mu ~, \nn\\
B_\mu &=& \left[1+\dfrac{{v_T^2}}{\Lambda^2}\,C_{HB}+\dfrac{{v_T^4}}{2\Lambda^4}\,C_{B^2H^4}^{(1)}
  +\dfrac{3{v_T^4}}{2\Lambda^4}\,[C_{HB}]^2\right]\mathcal B_\mu ~.
\eea

At this stage, there is still a $\mathcal W^3_{\mu\nu}\mathcal
B^{\mu\nu}$ mixing term, as well as a separate
$W^3_{\mu\nu}W^{3\mu\nu}$ term. These can both be removed by defining
\bea
\begin{bmatrix}
\mathcal W^3_\mu \\ \mathcal B_\mu
\end{bmatrix}=\begin{bmatrix}
1+\dfrac{{v_T}^4}{2\Lambda^4}\,C_{W^2H^4}^{(3)}+\dfrac{3{v_T}^4}{8\Lambda^4}\,[C_{HWB}]^2 & -\dfrac{{v_T}^2}{2\Lambda^2}\left(\begin{aligned}
& C_{HWB}+\dfrac{{v_T}^2}{2\Lambda^2}\,C_{WBH^4}^{(1)} \\
& +\dfrac{{v_T}^2}{\Lambda^2}\,C_{HWB}C_{HW} \\
& +\dfrac{{v_T}^2}{\Lambda^2}\,C_{HWB}C_{HB}
\end{aligned}\right) \\
-\dfrac{{v_T}^2}{2\Lambda^2}\left(\begin{aligned}
& C_{HWB}+\dfrac{{v_T}^2}{2\Lambda^2}\,C_{WBH^4}^{(1)} \\
& +\dfrac{{v_T}^2}{\Lambda^2}\,C_{HWB}C_{HW} \\
& +\dfrac{{v_T}^2}{\Lambda^2}\,C_{HWB}C_{HB}
\end{aligned}\right) & 1+\dfrac{3{v_T}^4}{8\Lambda^4}\,[C_{HWB}]^2
\end{bmatrix}\begin{bmatrix}
\overline{\mathcal W}^3_\mu \\ \overline{\mathcal B}_\mu
\end{bmatrix}.
\eea
With this, we have
\beq
\mathcal L_{\text{SMEFT}}^{\text{Electroweak kin}} = -\dfrac12\mathcal W^+_{\mu\nu}\mathcal W^{-\mu\nu}
-\dfrac14\overline{\mathcal W}^3_{\mu\nu}\overline{\mathcal W}^{3\mu\nu}
-\dfrac14\overline{\mathcal B}_{\mu\nu}\overline{\mathcal B}^{\mu\nu} ~,
\eeq
where $\mathcal W^\pm_{\mu\nu} \equiv \partial_\mu\mathcal W^\pm_\nu-\partial_\nu\mathcal W^\pm_\mu$, $\mathcal W^\pm_\mu \equiv \dfrac1{\sqrt2}(\mathcal W^1_\mu\mp i\mathcal W^2_\mu)$, $\overline{\mathcal W}^3_{\mu\nu} \equiv
\partial_\mu\overline{\mathcal W}^3_\nu-\partial_\nu\overline{\mathcal
  W}^3_\mu$, $\overline{\mathcal B}_{\mu\nu} \equiv
\partial_\mu\overline{\mathcal B}_\nu-\partial_\nu\overline{\mathcal
  B}_\mu$, and we have dropped the cubic and quartic self-coupling
terms of the gauge bosons.

Note that we still have the freedom to perform the following rotation:
\beq
\label{thetaWdef}
\begin{bmatrix}
\overline{\mathcal W}^3_\mu \\ \overline{\mathcal B}_\mu
\end{bmatrix}=\begin{bmatrix}
\cos\overline\theta_W & \sin\overline\theta_W \\
-\sin\overline\theta_W & \cos\overline\theta_W
\end{bmatrix}\begin{bmatrix}
\mathcal Z_\mu \\ \mathcal A_\mu
\end{bmatrix} ~,
\eeq
where $\mathcal Z_\mu$ and $\mathcal A_\mu$ are the physical $Z$-boson and photon fields. In terms of these fields, we have
\beq
\label{physicalGBs}
\mathcal L_{\text{SMEFT}}^{\text{Electroweak kin}}=-\dfrac12\mathcal W^+_{\mu\nu}\mathcal W^{-\mu\nu}
-\dfrac14\mathcal Z_{\mu\nu}\mathcal Z^{\mu\nu}-\dfrac14\mathcal F_{\mu\nu}\mathcal F^{\mu\nu} ~.
\eeq

For completeness, we also present the results for gluons. Including the
SMEFT contributions up to dimension 8, the gluon kinetic term is
\beq
\mathcal L_{\text{SMEFT}}^{\text{Gluons kin}}=-\dfrac14\left[1-\dfrac{2{v_T^2}}{\Lambda^2}\,C_{HG}-\dfrac{{v_T^4}}{\Lambda^4}\,C_{G^2H^4}^{(1)}\right]G^A_{\mu\nu}G^{A\mu\nu} ~.
\eeq
In order to properly normalize this kinetic term, we make
redefinitions similar to those in Eqs.~(\ref{ccredefs}) and
(\ref{gbredefs}):
\bea
\label{strong coupling constant dim 8}
\overline g_s &=& \left[1+\dfrac{{v_T^2}}{\Lambda^2}\,C_{HG}+\dfrac{{v_T^4}}{2\Lambda^4}\,C_{G^2H^4}^{(1)}
  +\dfrac{3{v_T^4}}{2\Lambda^4}\,[C_{HG}]^2\right]g_s ~, \\
\label{gluon field normalization dim 8}
G^A_\mu &=& \left[1+\dfrac{{v_T^2}}{\Lambda^2}\,C_{HG}+\dfrac{{v_T^4}}{2\Lambda^4}\,C_{G^2H^4}^{(1)}
  +\dfrac{3{v_T^4}}{2\Lambda^4}\,[C_{HG}]^2\right]\mathcal G^A_\mu ~.
\eea

\subsubsection{Mass terms}

The SMEFT contributions up to dimension 8 to the mass terms of the
electroweak gauge bosons after symmetry breaking are
\beq
\mathcal L_{\text{SMEFT}}^{\text{Electroweak mass}}=\dfrac{ v_T^2}8\left\{\begin{aligned}
& \left[1+\dfrac{ v_T^4}{4\Lambda^4}\left(C_{H^6}^{(1)}-C_{H^6}^{(2)}\right)\right]g^2(W_\mu^1W^{1\mu}+W_\mu^2W^{2\mu}) \\
& \left[\begin{aligned}
& 1+\dfrac{ v_T^2}{2\Lambda^2}\,C_{HD} \\
& +\dfrac{ v_T^4}{4\Lambda^4}\left(C_{H^6}^{(1)}+C_{H^6}^{(2)}\right)
\end{aligned}\right](gW_\mu^3-g'B_\mu)(gW^{3\mu}-g'B^\mu)
\end{aligned}\right\} ~.
\eeq
We can write $W_\mu^I$ and $B_\mu$ in terms of $\mathcal W^\pm_\mu$,
$\mathcal Z_\mu$ and $\mathcal A_\mu$ using the transformations
described in Sec.~\ref{GBkinetic}. The mixing angle ${\overline\theta}_W$ of Eq.~(\ref{thetaWdef}) satisfies
\begin{equation}
\label{mixing angle}
\begin{split}
\cos\overline\theta_W & =\dfrac1{\sqrt{\overline g^2+\overline g'^2}}\left[\begin{aligned}
& \overline g+\dfrac{\overline g{v_T^4}}{8\Lambda^4}\dfrac{(6\overline g^2\overline g'^2-\overline g^4-5\overline g'^4)}{(\overline g^2+\overline g'^2)^2}\,[C_{HWB}]^2+\dfrac{{v_T^4}}{2\Lambda^4}\dfrac{\overline g\overline g'^2}{\overline g^2+\overline g'^2}\,C_{W^2H^4}^{(3)} \\
& -\dfrac{\overline g'{v_T^2}}{2\Lambda^2}\dfrac{\overline g^2-\overline g'^2}{\overline g^2+\overline g'^2}\left(C_{HWB}+\dfrac{{v_T^2}}{2\Lambda^2}\,C_{WBH^4}^{(1)}+\dfrac{{v_T^2}}{\Lambda^2}\,C_{HWB}[C_{HW}+C_{HB}]\right)
\end{aligned}\right] \\
\sin\overline\theta_W & =\dfrac1{\sqrt{\overline g^2+\overline g'^2}}\left[\begin{aligned}
& \overline g'+\dfrac{\overline g'{v_T^4}}{8\Lambda^4}\dfrac{(6\overline g^2\overline g'^2-\overline g'^4-5\overline g^4)}{(\overline g^2+\overline g'^2)^2}\,[C_{HWB}]^2-\dfrac{{v_T^4}}{2\Lambda^4}\dfrac{\overline g^2\overline g'}{\overline g^2+\overline g'^2}\,C_{W^2H^4}^{(3)} \\
& +\dfrac{\overline g{v_T^2}}{2\Lambda^2}\dfrac{\overline g^2-\overline g'^2}{\overline g^2+\overline g'^2}\left(C_{HWB}+\dfrac{{v_T^2}}{2\Lambda^2}\,C_{WBH^4}^{(1)}+\dfrac{{v_T^2}}{\Lambda^2}\,C_{HWB}[C_{HW}+C_{HB}]\right)
\end{aligned}\right]
\end{split}
\end{equation}
up to dimension 8.

Note that, while in the SM we have $\sin\theta_W = g'/\sqrt{g^2 +
  {g'}^2}$ and $\cos\theta_W = g/\sqrt{g^2 + {g'}^2}$, these relations
no longer hold in the presence of SMEFT operators. Similarly, in the
SM, $e = gg'/\sqrt{g^2 + {g'}^2}$. Including SMEFT operators, this
becomes
\begin{equation}\label{electric charge}
\overline e=\dfrac{\overline g\overline g'}{\sqrt{\overline g^2+\overline g'^2}}\left[\begin{aligned}
& 1-\dfrac{\overline g\overline g' v_T^2C_{HWB}}{(\overline g^2+\overline g'^2)\Lambda^2}-\dfrac{\overline g\overline g' v_T^4C^{(1)}_{WBH^4}}{2(\overline g^2+\overline g'^2)\Lambda^4}+\dfrac{\overline g'^2 v_T^4C^{(3)}_{W^2H^4}}{2(\overline g^2+\overline g'^2)\Lambda^4} \\
& -\dfrac{\overline g\overline g' v_T^4C_{HWB}(C_{HW}+C_{HB})}{(\overline g^2+\overline g'^2)\Lambda^4}+\dfrac{3\overline g^2\overline g'^2 v_T^4[C_{HWB}]^2}{2(\overline g^2+\overline g'^2)^2\Lambda^4}
\end{aligned}\right] ~.
\end{equation}

The masses of the $W$ and $Z$ are given by
\bea
\label{W boson mass}
      M_W^2 &=& \dfrac{\overline g^2 v_T^2}4\left[1+\dfrac{ v_T^4}{4\Lambda^4}
        \left(C_{H^6}^{(1)}-C_{H^6}^{(2)}\right)\right] ~, \\
\label{Z boson mass}
M_Z^2 &=& \dfrac{{\overline g_Z}^2 v_T^2}4\left[1+\dfrac{ v_T^2}{2\Lambda^2}\,C_{HD}+\dfrac{ v_T^4}{4\Lambda^4}\left(C_{H^6}^{(1)}+C_{H^6}^{(2)}\right)\right] ~,
\eea
where
\bea
\label{Z coupling constant}
\overline g_Z &=& \sqrt{\overline g^2+\overline g'^2} \left[
  1 + \dfrac{\overline g\overline g' v_T^2C_{HWB}}{(\overline g^2+\overline g'^2)\Lambda^2}
  + \dfrac{\overline g\overline g' v_T^4C^{(1)}_{WBH^4}}{2(\overline g^2+\overline g'^2)\Lambda^4}
  + \dfrac{\overline g^2 v_T^4C^{(3)}_{W^2H^4}}{2(\overline g^2+\overline g'^2)\Lambda^4} \right. \nn\\
  && \left. + ~\dfrac{\overline g\overline g' v_T^4C_{HWB}(C_{HW}+C_{HB})}{(\overline g^2+\overline g'^2)\Lambda^4}
  + \left(1-\dfrac{\overline g^2\overline g'^2}{(\overline g^2+\overline g'^2)^2}\right)\dfrac{[C_{HWB}]^2}{2\Lambda^4}
\right] ~.
\eea

In the SM, the ``charge'' to which the $Z^0$ couples is $I_{3L} -
Q_{em} \sin^2 \theta_W$. When one adds SMEFT operators up to dimension
6, the mixing angle is changed, $\theta_W \to {\bar\theta_W}$, but the
$Z^0$ coupling still has the same form: it couples to $I_{3L} - Q_{em}
\sin^2 {\bar\theta_W}$ \cite{Jenkins:2017jig}. However, when SMEFT
operators up to dimension 8 are included, this no longer
holds. Instead, the $Z^0$ couples to $I_{3L} - Q_{em} \sin^2
{\bar\theta}_Z$, where
\begin{equation}
\label{new angle}
  \sin^2\overline\theta_Z=\sin^2\overline\theta_W
  +\dfrac{ v_T^4}{4\Lambda^4}[C_{HWB}]^2(\sin^2\overline\theta_W-\cos^2\overline\theta_W) ~.
\end{equation}
(This was also noted in Ref.~\cite{Hays:2018zze}.)

\subsection{Couplings of electroweak gauge bosons to fermions}

As shown in Eq.~(\ref{physicalGBs}), the physical electroweak gauge
bosons are $\mathcal A_\mu$, $\mathcal W_\mu^\pm$ and $\mathcal
Z_\mu$. Their effective couplings to fermions, as well as those of
the gluon $\mathcal G^A_\mu$, take the following form:
\begin{equation}
\label{fermion - gauge bosons interactions terms}
\mathcal L = -\overline g_s\mathcal G^A_\mu j_{\mathcal G}^{A\mu} -\overline e\mathcal A_\mu j_{\mathcal A}^\mu
-\dfrac{\overline g}{\sqrt2}\{\mathcal W^+_\mu j_{\mathcal W}^{+\mu}+\mathcal W^-_\mu j_{\mathcal W}^{-\mu}\}
-\overline g_Z\mathcal Z_\mu j_{\mathcal Z}^\mu ~,
\end{equation}
in which the corresponding currents are 
\bea
\label{fermion - gauge bosons currents}
j_{\mathcal G}^{A\mu} &=& \overline u_{Lp}\gamma^\mu T^Au_{Lr}+\overline d_{Lp}\gamma^\mu T^Ad_{Lr}
+\overline u_{Rp}\gamma^\mu T^Au_{Rr}+\overline d_{Rp}\gamma^\mu T^Ad_{Rr} ~, \nn\\
j_{\mathcal A}^\mu &=& -\overline e_{Lp}\gamma^\mu e_{Lr}+\dfrac23\overline u_{Lp}\gamma^\mu u_{Lr}
-\dfrac13\overline d_{Lp}\gamma^\mu d_{Lr}-\overline e_{Rp}\gamma^\mu e_{Rr}
+\dfrac23\overline u_{Rp}\gamma^\mu u_{Rr}-\dfrac13\overline d_{Rp}\gamma^\mu d_{Rr} ~, \nn\\
j_{\mathcal W}^{+\mu} &=& [W_l]^{\text{eff}}_{pr}\overline\nu_{Lp}\gamma^\mu e_{Lr}
+[W_q]^{\text{eff}}_{pr}\overline u_{Lp}\gamma^\mu d_{Lr}+[W_R]^{\text{eff}}_{pr}\overline u_{Rp}\gamma^\mu d_{Rr}+[W_l^{\slashed L}]^{\text{eff}}_{pr}(\nu_{Lp}^TC\gamma^\mu e_{Rr}) ~, \\
j_{\mathcal W}^{-\mu} &=& {[W_l]^{\text{eff}}_{rp}}^*\overline e_{Lp}\gamma^\mu\nu_{Lr}
+{[W_q]^{\text{eff}}_{rp}}^*\overline d_{Lp}\gamma^\mu u_{Lr}
+{[W_R]^{\text{eff}}_{rp}}^*\overline d_{Rp}\gamma^\mu u_{Rr}+{[W_l^{\slashed L}]^{\text{eff}}_{rp}}^*(\overline\nu_{Lp}\gamma^\mu C\overline e_{Rr}^T) ~, \nn\\
j_{\mathcal Z}^\mu &=& \left[\begin{aligned}
& [Z_{\nu_L}]^{\text{eff}}_{pr}\overline\nu_{Lp}\gamma^\mu\nu_{Lr}+[Z_{e_L}]^{\text{eff}}_{pr}\overline e_{Lp}\gamma^\mu e_{Lr}
+[Z_{u_L}]^{\text{eff}}_{pr}\overline u_{Lp}\gamma^\mu u_{Lr}+[Z_{d_L}]^{\text{eff}}_{pr}\overline d_{Lp}\gamma^\mu d_{Lr} \\
& +[Z_{e_R}]^{\text{eff}}_{pr}\overline e_{Rp}\gamma^\mu e_{Rr}+[Z_{u_R}]^{\text{eff}}_{pr}\overline u_{Rp}\gamma^\mu u_{Rr}+[Z_{d_R}]^{\text{eff}}_{pr}\overline d_{Rp}\gamma^\mu d_{Rr}
\end{aligned}\right] ~. \nn
\eea

Since $SU(3)_C \times U(1)_{em}$ remains unbroken, the currents involving gluons and photons are fully
determined by QCD and QED. This is not the case for the $\mathcal
W_\mu^\pm$ and $\mathcal Z_\mu$ gauge bosons: the fermion currents to which the $\mathcal
W_\mu^\pm$ and $\mathcal Z_\mu$ couple are given by the following (up
to dimension 8):
\begin{equation}
\label{gauge boson - fermions couplings}
\begin{split}
[W_l]^{\text{eff}}_{pr} & =\delta_{pr}+\dfrac{ v_T^2}{\Lambda^2}\,C_{\substack{Hl\\pr}}^{(3)}+\dfrac{ v_T^4}{2\Lambda^4}\left(C_{\substack{l^2H^4D\\pr}}^{(2)}-iC_{\substack{l^2H^4D\\pr}}^{(3)}+C_{\substack{l^2H^4D\\pr}}^{(4)}\right) ~, \\
[W_q]^{\text{eff}}_{pr} & =\delta_{pr}+\dfrac{ v_T^2}{\Lambda^2}\,C_{\substack{Hq\\pr}}^{(3)}+\dfrac{ v_T^4}{2\Lambda^4}\left(C_{\substack{q^2H^4D\\pr}}^{(2)}-iC_{\substack{q^2H^4D\\pr}}^{(3)}+C_{\substack{q^2H^4D\\pr}}^{(4)}\right) ~, \\
[W_R]^{\text{eff}}_{pr} & =\dfrac{ v_T^2}{2\Lambda^2}\,C_{\substack{Hud\\pr}}^{(3)}+\dfrac{ v_T^4}{4\Lambda^4}\,C_{\substack{udH^4D\\pr}} ~, \\
[W_l^{\slashed L}]^{\text{eff}}_{pr} & =-\dfrac{{v_T}^3}{2\sqrt2\,\Lambda^3}C_{\substack{leH^3D\\pr}} ~, \\
[Z_{\nu_L}]^{\text{eff}}_{pr} & =\dfrac12\delta_{pr}-\dfrac{ v_T^2}{2\Lambda^2}\left(C_{\substack{Hl\\pr}}^{(1)}-C_{\substack{Hl\\pr}}^{(3)}\right)-\dfrac{ v_T^4}{4\Lambda^4}\left(C_{\substack{l^2H^4D\\pr}}^{(1)}-2C_{\substack{l^2H^4D\\pr}}^{(2)}\right) ~, \\
[Z_{e_L}]^{\text{eff}}_{pr} & =\dfrac12g^e_L\delta_{pr}-\dfrac{ v_T^2}{2\Lambda^2}\left(C_{\substack{Hl\\pr}}^{(1)}+C_{\substack{Hl\\pr}}^{(3)}\right)-\dfrac{ v_T^4}{4\Lambda^4}\left(C_{\substack{l^2H^4D\\pr}}^{(1)}+2C_{\substack{l^2H^4D\\pr}}^{(2)}\right) ~, \\
[Z_{u_L}]^{\text{eff}}_{pr} & =\dfrac12g^u_L\delta_{pr}-\dfrac{ v_T^2}{2\Lambda^2}\left(C_{\substack{Hq\\pr}}^{(1)}-C_{\substack{Hq\\pr}}^{(3)}\right)-\dfrac{ v_T^4}{4\Lambda^4}\left(C_{\substack{q^2H^4D\\pr}}^{(1)}-2C_{\substack{q^2H^4D\\pr}}^{(2)}\right) ~, \\
[Z_{d_L}]^{\text{eff}}_{pr} & =\dfrac12g^d_L\delta_{pr}-\dfrac{ v_T^2}{2\Lambda^2}\left(C_{\substack{Hq\\pr}}^{(1)}+C_{\substack{Hq\\pr}}^{(3)}\right)-\dfrac{ v_T^4}{4\Lambda^4}\left(C_{\substack{q^2H^4D\\pr}}^{(1)}+2C_{\substack{q^2H^4D\\pr}}^{(2)}\right) ~, \\
[Z_{e_R}]^{\text{eff}}_{pr} & =\dfrac12g^e_R\delta_{pr}-\dfrac{ v_T^2}{2\Lambda^2}\,C_{\substack{He\\pr}}-\dfrac{ v_T^4}{4\Lambda^4}\,C_{\substack{e^2H^4D\\pr}}^{(1)} ~, \\
[Z_{u_R}]^{\text{eff}}_{pr} & =\dfrac12g^u_R\delta_{pr}-\dfrac{v^2}{2\Lambda^2}\,C_{\substack{Hu\\pr}}-\dfrac{ v_T^4}{4\Lambda^4}\,C_{\substack{u^2H^4D\\pr}}^{(1)} ~, \\
[Z_{d_R}]^{\text{eff}}_{pr} & =\dfrac12g^d_R\delta_{pr}-\dfrac{v^2}{2\Lambda^2}\,C_{\substack{Hd\\pr}}-\dfrac{ v_T^4}{4\Lambda^4}\,C_{\substack{d^2H^4D\\pr}}^{(1)} ~.
\end{split}
\end{equation}
Here, we have defined $g^e_L\equiv-1+2\sin^2\overline\theta_Z$,
$g^u_L\equiv1-\dfrac43\sin^2\overline\theta_Z$,
$g^d_L\equiv-1+\dfrac23\sin^2\overline\theta_Z$,
$g^e_R\equiv2\sin^2\overline\theta_Z$,
$g^u_R\equiv-\dfrac43\sin^2\overline\theta_Z$, and
$g^d_R\equiv\dfrac23\sin^2\overline\theta_Z$, where
$\sin^2\overline\theta_Z$ is defined in Eq.~(\ref{new angle}).

\section{Matching conditions}

There are four classes of LEFT operators up to dimension 6: (i) four-fermion operators, (ii) magnetic dipole moment operators, (iii) three-gluon operators, and (iv) neutrino mass terms. The matching conditions for operators that conserve both $B$ and $L$ involve only even-dimension SMEFT operators, and are given up to dimension 6 in JMS. For operators that violate $B$ and/or $L$, the matching conditions involve either even- or odd-dimension SMEFT operators, depending on the operator; these are given to dimension 6 or dimension 5 in JMS. 

{In general, there are three types of contributions to the matching conditions of LEFT operators to SMEFT operators up to dimension-8: (a) \emph{direct} contributions, (b) \emph{indirect} contributions, and (c) \emph{momentum-dependent} contributions. The classes (ii)-(iv) of LEFT operators receive only direct contributions.  In the following subsections, we describe in detail the direct and indirect contributions to four-fermion operators and their origin within the SMEFT up to the dimension-8 level. We also briefly summarize the sources of the MD contributions.}

\subsection{Direct contributions}

The dimension-6 SMEFT direct contribution is the LEFT operator itself, in
which all left- and right-handed particles are replaced by the left-handed
$SU(2)_L$ doublets and right-handed $SU(2)_L$ singlets to which they respectively belong.  The dimension-8
contributions involve the dimension-6 SMEFT operator multiplied by a
pair of Higgs fields. When the Higgs gets a vev, this generates the
four-fermion LEFT operator.

The details of the computation are best illustrated with an example.
Consider the LEFT operator
\beq
\mathcal O^{V,LL}_{\substack{\nu\nu\\prst}} \equiv (\overline\nu_{Lp}
\gamma_\mu \nu_{Lr}) (\overline\nu_{Ls} \gamma^\mu \nu_{Lt}) ~.
\label{LEFT4nu}
\eeq
It is generated by
the dimension-6 SMEFT operator $Q_{\substack{ll\\prst}} \equiv (\overline
l_p\gamma_\mu l_r)(\overline l_s\gamma^\mu l_t)$. This can be seen by
separating the SMEFT operator into components:
\beq
\dfrac1{\Lambda^2} C_{\substack{ll\\prst}} Q_{\substack{ll\\prst}} \to\dfrac1{\Lambda^2}
C_{\substack{ll\\prst}}\left[\begin{aligned}
& (\overline\nu_{Lp}\gamma_\mu\nu_{Lr})(\overline\nu_{Ls}\gamma^\mu\nu_{Lt})+(\overline\nu_{Lp}\gamma_\mu\nu_{Lr})(\overline e_{Ls}\gamma^\mu e_{Lt}) \\
& +(\overline e_{Lp}\gamma_\mu e_{Lr})(\overline\nu_{Ls}\gamma^\mu\nu_{Lt})+(\overline e_{Lp}\gamma_\mu e_{Lr})(\overline e_{Ls}\gamma^\mu e_{Lt})
\end{aligned}\right] ~.
\eeq
The first term is $\mathcal O^{V,LL}_{\substack{\nu\nu\\prst}}$.

One dimension-8 SMEFT operator that is among the matching conditions
is $Q^{(1)}_{\substack{l^4H^2\\prst}} \equiv (\overline l_p\gamma_\mu l_r)
(\overline l_s\gamma^\mu l_t)(H^\dagger H)$. Because the $SU(2)_L$
doublets $l$ and $H$ are involved, there are two additional
dimension-8 SMEFT operators that must be included:
$Q^{(2)}_{\substack{l^4H^2\\prst}}\equiv(\overline l_p\gamma_\mu
l_r)(\overline l_s\gamma^\mu\tau^Il_t)(H^\dagger\tau^IH)$ and
$Q^{(2)}_{\substack{l^4H^2\\stpr}}\equiv(\overline
l_p\gamma_\mu\tau^Il_r)(\overline l_s\gamma^\mu
l_t)(H^\dagger\tau^IH)$. When the Higgs gets a vev, these three
operators can also generate $\mathcal
O^{V,LL}_{\substack{\nu\nu\\prst}}$:
\bea
\dfrac1{\Lambda^4}C^{(1)}_{\substack{l^4H^2\\prst}}Q^{(1)}_{\substack{l^4H^2\\prst}} &\to&
\dfrac{{v_T^2}}{2\Lambda^4}C^{(1)}_{\substack{l^4H^2\\prst}}\left[\begin{aligned}
& (\overline\nu_{Lp}\gamma_\mu\nu_{Lr})(\overline\nu_{Ls}\gamma^\mu\nu_{Lt})+(\overline\nu_{Lp}\gamma_\mu\nu_{Lr})(\overline e_{Ls}\gamma^\mu e_{Lt}) \\
& +(\overline e_{Lp}\gamma_\mu e_{Lr})(\overline\nu_{Ls}\gamma^\mu\nu_{Lt})+(\overline e_{Lp}\gamma_\mu e_{Lr})(\overline e_{Ls}\gamma^\mu e_{Lt})
\end{aligned}\right] ~, \nn\\
\dfrac1{\Lambda^4}C^{(2)}_{\substack{l^4H^2\\prst}}Q^{(2)}_{\substack{l^4H^2\\prst}} &\to&
\dfrac{{v_T^2}}{2\Lambda^4}C^{(2)}_{\substack{l^4H^2\\prst}}\left[\begin{aligned}
& -(\overline\nu_{Lp}\gamma_\mu\nu_{Lr})(\overline\nu_{Ls}\gamma^\mu\nu_{Lt})+(\overline\nu_{Lp}\gamma_\mu\nu_{Lr})(\overline e_{Ls}\gamma^\mu e_{Lt}) \\
& -(\overline e_{Lp}\gamma_\mu e_{Lr})(\overline\nu_{Ls}\gamma^\mu\nu_{Lt})+(\overline e_{Lp}\gamma_\mu e_{Lr})(\overline e_{Ls}\gamma^\mu e_{Lt})
\end{aligned}\right] ~, \nn\\
\dfrac1{\Lambda^4}C^{(2)}_{\substack{l^4H^2\\stpr}}Q^{(2)}_{\substack{l^4H^2\\stpr}} &\to&
\dfrac{{v_T^2}}{2\Lambda^4}C^{(2)}_{\substack{l^4H^2\\stpr}}\left[\begin{aligned}
& -(\overline\nu_{Lp}\gamma_\mu\nu_{Lr})(\overline\nu_{Ls}\gamma^\mu\nu_{Lt})-(\overline\nu_{Lp}\gamma_\mu\nu_{Lr})(\overline e_{Ls}\gamma^\mu e_{Lt}) \\
& +(\overline e_{Lp}\gamma_\mu e_{Lr})(\overline\nu_{Ls}\gamma^\mu\nu_{Lt})+(\overline e_{Lp}\gamma_\mu e_{Lr})(\overline e_{Ls}\gamma^\mu e_{Lt})
  \end{aligned}\right] ~.
\eea

We therefore see that the direct contribution to the matching
condition of the LEFT operator $\dfrac1{\Lambda^2}\mathcal
O^{V,LL}_{\substack{\nu\nu\\prst}}$, up to dimension 8, is
\beq
\dfrac1{\Lambda^2}\left[C_{\substack{ll\\prst}}+\dfrac{{v_T^2}}{2\Lambda^2}
  \left(C^{(1)}_{\substack{l^4H^2\\prst}}-C^{(2)}_{\substack{l^4H^2\\prst}}-C^{(2)}_{\substack{l^4H^2\\stpr}}\right)\right] ~.
\eeq

The direct contributions to the matching conditions of the other LEFT
four-fermion operators are calculated similarly.

\subsection{Indirect contributions}

A four-fermion operator can also be generated when a boson is
exchanged between two fermion currents and this boson is integrated
out. This produces an indirect contribution [e.g., see
  Eq.~(\ref{SMmatch})].

\begin{figure}[!htbp]
\begin{center}
\includegraphics[width=0.3\textwidth]{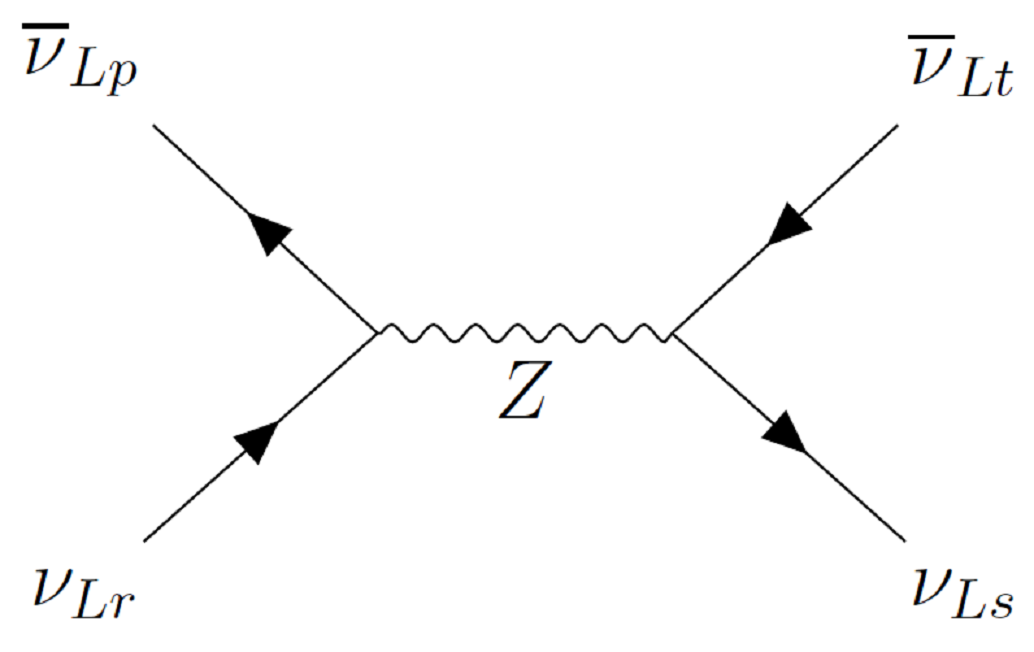}
~~~~~~~
\includegraphics[width=0.3\textwidth]{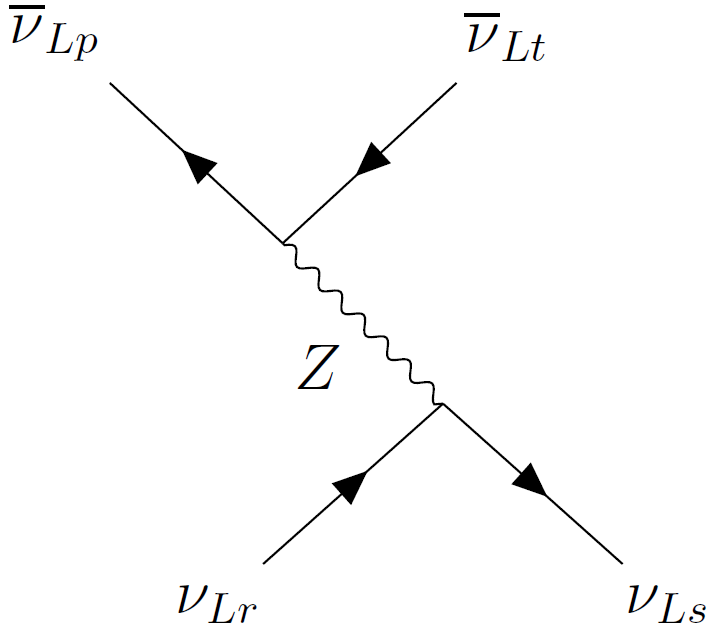}
\end{center}
\vskip-5truemm
\caption{\small $Z$-exchange contributions to $\mathcal O^{V,LL}_{\nu\nu}$ with flavour indices $prst$.}
\label{4nuZexchange}
\end{figure}

Consider once again the LEFT operator $\mathcal
O^{V,LL}_{\substack{\nu\nu\\prst}}$ of Eq.~(\ref{LEFT4nu}). The
indirect contributions arise from the $Z$-exchange diagrams of
Fig.~\ref{4nuZexchange}, when the $Z^0$ is integrated out. We note
that (i) there is a relative minus sign between the two diagrams, and
(ii) when one Fierz transforms (see Appendix \ref{Fierz}) the amplitude of the second diagram,
one obtains the amplitude of the first diagram, but with an exchange
of generation indices $r \leftrightarrow t$. The indirect contribution
to the matching condition of this operator, up to dimension 8, is
\beq
-\dfrac{\overline{g}_Z^2}{4 M_Z^2} \left([Z_{\nu_L}]^{\text{eff}}_{pr}[Z_{\nu_L}]^{\text{eff}}_{st}
+[Z_{\nu_L}]^{\text{eff}}_{pt}[Z_{\nu_L}]^{\text{eff}}_{sr}\right) ~,
\eeq
where $\overline{g}_Z$ and $[Z_{\nu_L}]^{\text{eff}}_{pr}$ are defined
in Eqs.~(\ref{Z coupling constant}) and (\ref{gauge boson - fermions
  couplings}), respectively.

\begin{figure}[!htbp]
\begin{center}
\includegraphics[width=0.3\textwidth]{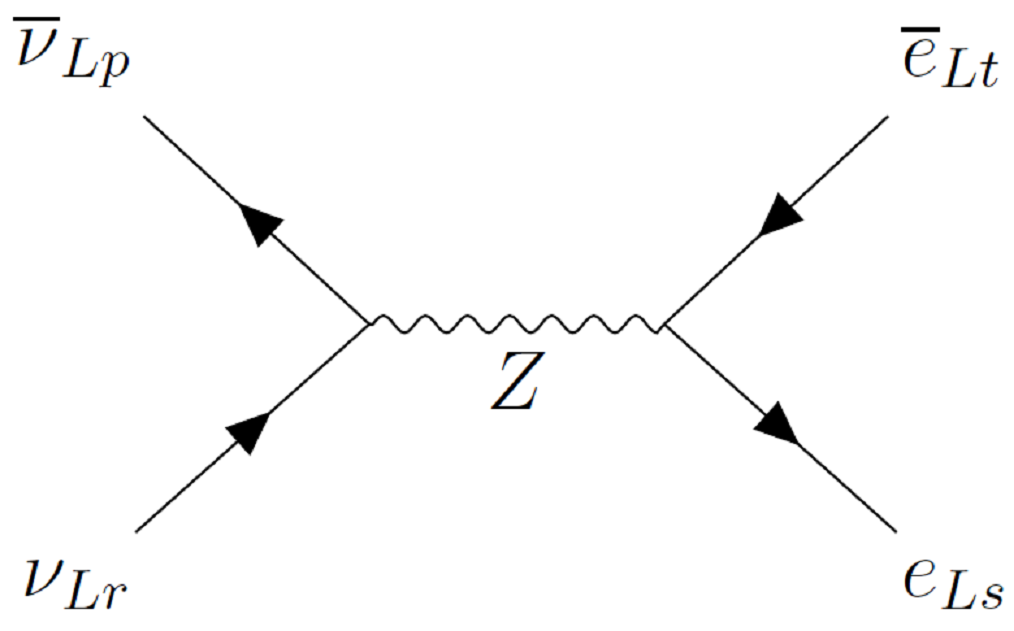}
~~~~~~~
\includegraphics[width=0.3\textwidth]{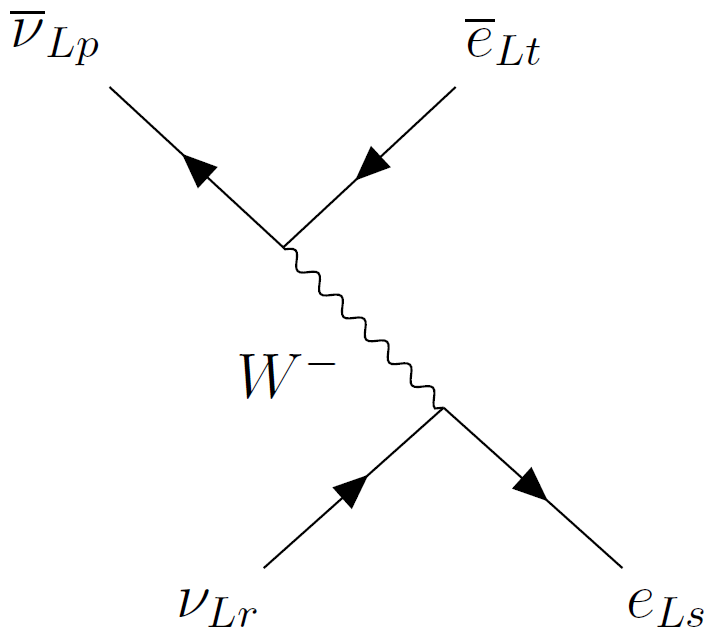}
\end{center}
\vskip-5truemm
\caption{\small $Z$- and $W$-exchange contributions to $\mathcal O^{V,LL}_{\nu e}$ with flavour indices $prst$.}
\label{2nu2eZWexchange}
\end{figure}

Another example is the LEFT operator $\mathcal O^{V,LL}_{\substack{\nu
    e\\prst}} \equiv (\overline\nu_{Lp}\gamma_\mu\nu_{Lr})(\overline
e_{Ls}\gamma^\mu e_{Lt})$. Here the indirect contributions arise from
the $Z$- and $W$-exchange diagrams of Fig,~\ref{2nu2eZWexchange}, when
the heavy gauge bosons are integrated out. The indirect contribution
to the matching condition of this operator, up to dimension 8, is
\beq
-\dfrac{\overline{g}_Z^2}{M_Z^2}[Z_{\nu_L}]^{\text{eff}}_{pr}[Z_{e_L}]^{\text{eff}}_{st}
-\dfrac{\overline g^2}{2 M_W^2}[W_l]^{\text{eff}}_{pr}{[W_l]^{\text{eff}}_{st}}^* ~,
\eeq
where $\overline{g}$ and $[W_l]^{\text{eff}}_{pr}$ are defined in
Eqs.~(\ref{ccredefs}) and (\ref{gauge boson - fermions couplings}),
respectively.

The indirect contributions from gauge-boson exchange to the matching
conditions of the other LEFT four-fermion operators are calculated
similarly. Most such operators can be generated via diagrams with the
exchange of a $Z^0$. A small subset of these also involve $W$-exchange
diagrams. And a few LEFT operators can be generated only via the
exchange of a $W$.

Finally, the matching conditions of certain LEFT operators receive
indirect contributions from Higgs exchange. As an example, consider
the operator $\mathcal O^{V,LR}_{\substack{ee\\prst}} \equiv
(\overline e_{Lp}\gamma_\mu e_{Lr})(\overline e_{Rs}\gamma^\mu
e_{Rt})$. 

\begin{figure}[!htbp]
\begin{center}
\includegraphics[width=0.3\textwidth]{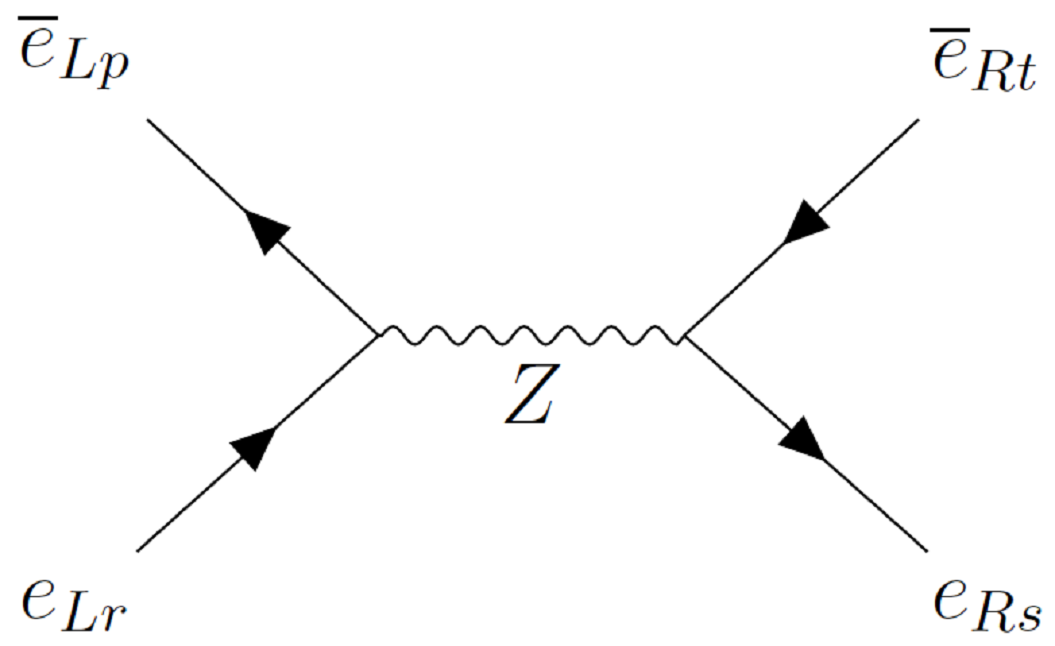}
~~~~~~~
\includegraphics[width=0.3\textwidth]{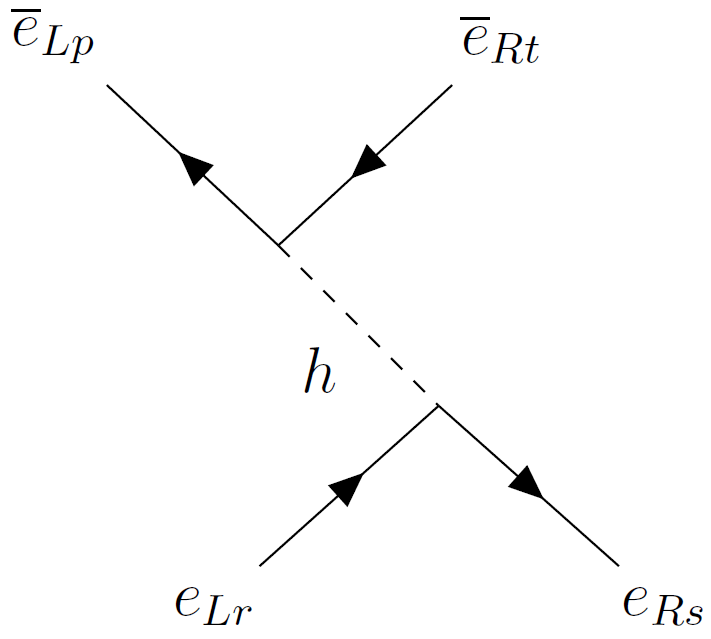}
\end{center}
\vskip-5truemm
\caption{\small $Z$- and $h$-exchange contributions to $\mathcal O^{V,LL}_{\nu e}$ with flavour indices $prst$.}
\label{eLeRZhexchange}
\end{figure}

The indirect contributions come from the diagrams of
Fig.~\ref{eLeRZhexchange}. The $Z$- and $h$-exchange contributions are
computed similarly to the previous examples. The indirect contribution
to the matching condition is
\beq
-\dfrac{\overline{g_Z}^2}{{M_Z}^2}[Z_{e_L}]^{\text{eff}}_{pr}[Z_{e_R}]^{\text{eff}}_{st}
-\dfrac1{2{m_h}^2}\,(Y_e)^{\text{eff}}_{pt}{(Y_e)^{\text{eff}}_{rs}}^* ~.
\eeq
The Yukawa coupling is [Eq.~(\ref{Yukawa couplings dimension 8}),
  repeated for convenience]
\[
(Y_e)^{\text{eff}}_{pr} = \dfrac{1+c_{H,\text{kin}}}{\sqrt2}\left[\dfrac{\sqrt2}{v_T}\,[M_e]_{pr}-\dfrac{{v_T^2}}{\Lambda^2}\,C_{\substack{e H\\pr}}-\dfrac{{v_T^4}}{\Lambda^4}\,C_{\substack{\chi e H^5\\pr}}\right] ~.
\]
The first term is $\sim m_e/v_T$ and is negligible. For this reason,
JMS, which works only to dimension 6, argues that the $h$-exchange
indirect contributions to the matching conditions are unimportant.
However, when one works to dimension 8, there is a non-negligible
contribution resulting from the square of the second term.

\subsection{Momentum-dependent contributions}
\label{MDcont}

{In the introduction, we noted that (i) the matching conditions can be separated into two types, momentum-independent and momentum-dependent, and (ii) in this paper, we focus only on the MI type. Indeed, in the above subsections, the direct and indirect contributions give rise to MI matching conditions. Still, it is a useful exercise to explore which types of SMEFT operators can produce MD matching conditions. The various sources of MD contributions are outlined below.}

{We first consider dimension-8 SMEFT operators. Those belonging to the $\psi^4 D^2$ class contribute directly to four-fermion operators and give rise to MD contributions that scale as $p^2/\Lambda^4$. There are also MD operators in the $\psi^2 H^4 D$ class. These contribute to four-fermion operators via Higgs exchange; the net effect scales as $v p / \Lambda^4$. However, the MI contributions scale as $v^2/\Lambda^4$.  Since the momentum transfer $p$ in low-energy processes is much smaller than $v$, one can safely neglect these types of MD contributions.}

{On the other hand, there are also MD contributions from dimension-6 SMEFT operators. Those in the $\psi^2 H^2 D$ and $\psi^2 X H$ classes contribute to four-fermion operators via the exchange of a Higgs boson or a $W$/$Z$ boson, respectively. These contributions scale as 
$(m/v) (1/v^2) (p v/\Lambda^2) \sim (p^2/v^2) /\Lambda^2$ and $(1/v^2) (p v/\Lambda^2) \sim (p/v) /\Lambda^2$, respectively. Clearly, depending on the values of $\Lambda$ and $p \sim m$, these can be comparable to the MI dimension-8 SMEFT contributions, which scale as $v^2/\Lambda^4$.}

{In addition, the second-order term in the expansion of the propagator can give rise to contributions of the same order, $(p^2/v^4) (v^2/\Lambda^2) \sim (p^2/v^2)/\Lambda^2$ . In this case, the momentum dependence arises from the propagator, in contrast to the above contributions, where it is in the vertex. (Note that, with dimension-8 operators,  this type of effect is suppressed since it scales as $(p^2/v^4) (v^2/\Lambda^4) \sim (p^2/v^2)/\Lambda^4$, which can be neglected.}

\subsection{Results}

For all four-fermion LEFT operators, the {MI} matching conditions  up to dimension 8 in SMEFT are determined using the techniques described above for computing the direct and indirect contributions. For the LEFT magnetic dipole moment operators, three-gluon operators and neutrino mass terms, the calculations are straightforward, as there are no indirect contributions. The matching conditions are given in the tables in Appendix \ref{Matchings}.

In the literature, the matching conditions of LEFT operators to dimension-7 SMEFT operators have been calculated in Ref.~\cite{XiaoDongMaMatchings}. The results obtained there are in agreement with ours. 
The matching conditions of LEFT operators to dimension-8 SMEFT operators has only been performed in Refs.~\cite{Ardu:2021koz, Ardu:2022pzk}, where the focus was on LEFT operators that lead to lepton flavour violation. Our results agree with this analysis. Matching conditions to dimension-8 SMEFT operators have also been computed in Ref.~\cite{Boughezal:2021tih}, but in the context of high-energy processes. Although LEFT operators were not involved, there is still some overlap, and we agree here as well. Finally, the contributions of dimension-8 SMEFT operators to the SM parameters, as described in Sec.~\ref{Setup}, was also examined in Ref.~\cite{Ardu:2021koz}, and we are in agreement.

\section{Conclusions}

The modern thinking is that the Standard Model is the leading part of an effective field theory, produced when the heavy new physics is integrated out. This EFT is usually assumed to be the SMEFT, which includes the Higgs boson. The SMEFT has been well-studied – all operators up to dimension 8 have been worked out. 

When the heavy particles of the SM ($W^\pm$, $Z^0$, $H$, $t$) are also integrated out, one obtains the LEFT (low-energy EFT), applicable at scales {$\ll M_W$}. In order to establish how low-energy measurements are affected by the underlying NP, it is necessary to determine how the LEFT operators depend on the SMEFT operators (the matching conditions). 

In Ref.~\cite{Jenkins:2017jig}, Jenkins, Manohar and Stoffer (JMS) present a complete and non-redundant basis of LEFT operators up to dimension 6, and compute the matching to SMEFT operators up to dimension 6. However, if the low-energy observable in question is suppressed in the SM and/or is very precisely measured, this may not be sufficient. Indeed, it has been pointed out that dimension-8 SMEFT contributions may be important for electroweak precision data from LEP, lepton-flavour-violating processes, meson-antimeson mixing, and electric dipole moments. 

In this paper, we extend the analysis of JMS: for all LEFT operators, we work out the complete {tree-level momentum-independent} matching conditions to SMEFT operators up to dimension 8. {The momentum-dependent contributions will be presented elsewhere.} There are direct contributions to these matching conditions for all LEFT operators, and four-fermion operators also receive indirect contributions due to the exchange of a $W^\pm$, $Z^0$ and/or $H$. 

Should the analysis of a LEFT observable require information about dimension-8 SMEFT {tree-level contributions, much of} that information can be found here.

\section*{Acknowledgements}
We thank many people for helpful discussions about their work: M. Ardu
and S. Davidson (2103.07212), C.W. Murphy (2005.00059), E.E. Jenkins,
A.V. Manohar and P. Stoffer (1709.04486), C. Hays, A. Martin, V. Sanz
and J. Setford (1808.00442), R. Boughezal, E. Mereghetti and
F. Petriello (2106.05337). We also thank Xiao-Dong Ma for helpful discussions, for bringing 
Refs.~\cite{XiaoDongMaMatchings, XiaoDongMa} to our attention, and for correcting an error in the matching conditions involving some of the fermion-number-violating SMEFT operators. We also thank the authors of \cite{UFO} for pointing out that the previous version of our paper was missing contributions from the mass-dimension $8$ operators $Q^{(4)}_{\substack{l^2H^4D\\pr}}:\epsilon^{IJK}(\overline l_p\gamma^\mu\tau^Il_r)(H^\dagger\tau^JH)D_\mu(H^\dagger\tau^KH)$ and $Q^{(4)}_{\substack{q^2H^4D\\pr}}:\epsilon^{IJK}(\overline q_p\gamma^\mu\tau^Iq_r)(H^\dagger\tau^JH)D_\mu(H^\dagger\tau^KH)$ to the effective SMEFT $W$-lepton and $W$-quark effective couplings, which are described in the first two of equations (\ref{gauge boson - fermions couplings}). This work was partially financially
supported by NSERC of Canada (SH, DL). JK is financially supported by
a postdoctoral research fellowship of the Alexander von Humboldt
Foundation.

\appendix

\section{LEFT operators up to dimension 6}

The following two tables are taken from Ref.~\cite{Jenkins:2017jig}.
\begin{table}[H]
\vspace{-0.75cm}
\begin{adjustbox}{width=0.85\textwidth,center}
\begin{minipage}[t]{3cm}
\renewcommand{\arraystretch}{1.51}
\small
\begin{align*}
\begin{array}[t]{c|c}
\multicolumn{2}{c}{\boldsymbol{\nu \nu+\hc}} \\
\hline
\O_{\nu} & (\nu_{Lp}^T C \nu_{Lr})  \\
\end{array}
\end{align*}
\end{minipage}
% --- END TABLE
%%
%
%%
% --- START TABLE
%%
\begin{minipage}[t]{3cm}
\renewcommand{\arraystretch}{1.51}
\small
\begin{align*}
\begin{array}[t]{c|c}
\multicolumn{2}{c}{\boldsymbol{(\nu \nu) X+\hc}} \\
\hline
\O_{\nu \gamma} & (\nu_{Lp}^T C   \sigma^{\mu \nu}  \nu_{Lr})  F_{\mu \nu}  \\
\end{array}
\end{align*}
\end{minipage}
%%%
% --- END TABLE
%%%
%%
% --- START TABLE
%%
\begin{minipage}[t]{3cm}
\renewcommand{\arraystretch}{1.51}
\small
\begin{align*}
\begin{array}[t]{c|c}
\multicolumn{2}{c}{\boldsymbol{(\overline L R ) X+\hc}} \\
\hline
\O_{e \gamma} & \bar e_{Lp}   \sigma^{\mu \nu} e_{Rr}\, F_{\mu \nu}  \\
\O_{u \gamma} & \bar u_{Lp}   \sigma^{\mu \nu}  u_{Rr}\, F_{\mu \nu}   \\
\O_{d \gamma} & \bar d_{Lp}  \sigma^{\mu \nu} d_{Rr}\, F_{\mu \nu}  \\
\O_{u G} & \bar u_{Lp}   \sigma^{\mu \nu}  T^A u_{Rr}\,  G_{\mu \nu}^A  \\
\O_{d G} & \bar d_{Lp}   \sigma^{\mu \nu} T^A d_{Rr}\,  G_{\mu \nu}^A \\
\end{array}
\end{align*}
\end{minipage}
\begin{minipage}[t]{3cm}
\renewcommand{\arraystretch}{1.51}
\small
\begin{align*}
\begin{array}[t]{c|c}
\multicolumn{2}{c}{\boldsymbol{X^3}} \\
\hline
\O_G     & f^{ABC} G_\mu^{A\nu} G_\nu^{B\rho} G_\rho^{C\mu}  \\
\O_{\widetilde G} & f^{ABC} \widetilde G_\mu^{A\nu} G_\nu^{B\rho} G_\rho^{C\mu}   \\
\end{array}
\end{align*}
\end{minipage}
\end{adjustbox}
%

%
%%
% --- START TABLE
%%
\mbox{}\\[-1.25cm]

\begin{adjustbox}{width=1.05\textwidth,center}
%%%
% --- END TABLE
%%%
\begin{minipage}[t]{3cm}
\renewcommand{\arraystretch}{1.51}
\small
\begin{align*}
\begin{array}[t]{c|c}
\multicolumn{2}{c}{\boldsymbol{(\overline L L)(\overline L  L)}} \\
\hline
\op{\nu\nu}{V}{LL} & (\bar \nu_{Lp}  \gamma^\mu \nu_{Lr} )(\bar \nu_{Ls} \gamma_\mu \nu_{Lt})   \\
\op{ee}{V}{LL}       & (\bar e_{Lp}  \gamma^\mu e_{Lr})(\bar e_{Ls} \gamma_\mu e_{Lt})   \\
\op{\nu e}{V}{LL}       & (\bar \nu_{Lp} \gamma^\mu \nu_{Lr})(\bar e_{Ls}  \gamma_\mu e_{Lt})  \\
\op{\nu u}{V}{LL}       & (\bar \nu_{Lp} \gamma^\mu \nu_{Lr}) (\bar u_{Ls}  \gamma_\mu u_{Lt})  \\
\op{\nu d}{V}{LL}       & (\bar \nu_{Lp} \gamma^\mu \nu_{Lr})(\bar d_{Ls} \gamma_\mu d_{Lt})     \\
\op{eu}{V}{LL}      & (\bar e_{Lp}  \gamma^\mu e_{Lr})(\bar u_{Ls} \gamma_\mu u_{Lt})   \\
\op{ed}{V}{LL}       & (\bar e_{Lp}  \gamma^\mu e_{Lr})(\bar d_{Ls} \gamma_\mu d_{Lt})  \\
\op{\nu edu}{V}{LL}      & (\bar \nu_{Lp} \gamma^\mu e_{Lr}) (\bar d_{Ls} \gamma_\mu u_{Lt})  + \hc   \\
\op{uu}{V}{LL}        & (\bar u_{Lp} \gamma^\mu u_{Lr})(\bar u_{Ls} \gamma_\mu u_{Lt})    \\
\op{dd}{V}{LL}   & (\bar d_{Lp} \gamma^\mu d_{Lr})(\bar d_{Ls} \gamma_\mu d_{Lt})    \\
\op{ud}{V1}{LL}     & (\bar u_{Lp} \gamma^\mu u_{Lr}) (\bar d_{Ls} \gamma_\mu d_{Lt})  \\
\op{ud}{V8}{LL}     & (\bar u_{Lp} \gamma^\mu T^A u_{Lr}) (\bar d_{Ls} \gamma_\mu T^A d_{Lt})   \\[-0.5cm]
\end{array}
\end{align*}
\renewcommand{\arraystretch}{1.51}
\small
\begin{align*}
\begin{array}[t]{c|c}
\multicolumn{2}{c}{\boldsymbol{(\overline R  R)(\overline R R)}} \\
\hline
\op{ee}{V}{RR}     & (\bar e_{Rp} \gamma^\mu e_{Rr})(\bar e_{Rs} \gamma_\mu e_{Rt})  \\
\op{eu}{V}{RR}       & (\bar e_{Rp}  \gamma^\mu e_{Rr})(\bar u_{Rs} \gamma_\mu u_{Rt})   \\
\op{ed}{V}{RR}     & (\bar e_{Rp} \gamma^\mu e_{Rr})  (\bar d_{Rs} \gamma_\mu d_{Rt})   \\
\op{uu}{V}{RR}      & (\bar u_{Rp} \gamma^\mu u_{Rr})(\bar u_{Rs} \gamma_\mu u_{Rt})  \\
\op{dd}{V}{RR}      & (\bar d_{Rp} \gamma^\mu d_{Rr})(\bar d_{Rs} \gamma_\mu d_{Rt})    \\
\op{ud}{V1}{RR}       & (\bar u_{Rp} \gamma^\mu u_{Rr}) (\bar d_{Rs} \gamma_\mu d_{Rt})  \\
\op{ud}{V8}{RR}    & (\bar u_{Rp} \gamma^\mu T^A u_{Rr}) (\bar d_{Rs} \gamma_\mu T^A d_{Rt})  \\
\end{array}
\end{align*}
\end{minipage}
%
%\hspace{-1.5cm}
%
\begin{minipage}[t]{3cm}
\renewcommand{\arraystretch}{1.51}
\small
\begin{align*}
\begin{array}[t]{c|c}
\multicolumn{2}{c}{\boldsymbol{(\overline L  L)(\overline R  R)}} \\
\hline
\op{\nu e}{V}{LR}     & (\bar \nu_{Lp} \gamma^\mu \nu_{Lr})(\bar e_{Rs}  \gamma_\mu e_{Rt})  \\
\op{ee}{V}{LR}       & (\bar e_{Lp}  \gamma^\mu e_{Lr})(\bar e_{Rs} \gamma_\mu e_{Rt}) \\
\op{\nu u}{V}{LR}         & (\bar \nu_{Lp} \gamma^\mu \nu_{Lr})(\bar u_{Rs}  \gamma_\mu u_{Rt})    \\
\op{\nu d}{V}{LR}         & (\bar \nu_{Lp} \gamma^\mu \nu_{Lr})(\bar d_{Rs} \gamma_\mu d_{Rt})   \\
\op{eu}{V}{LR}        & (\bar e_{Lp}  \gamma^\mu e_{Lr})(\bar u_{Rs} \gamma_\mu u_{Rt})   \\
\op{ed}{V}{LR}        & (\bar e_{Lp}  \gamma^\mu e_{Lr})(\bar d_{Rs} \gamma_\mu d_{Rt})   \\
\op{ue}{V}{LR}        & (\bar u_{Lp} \gamma^\mu u_{Lr})(\bar e_{Rs}  \gamma_\mu e_{Rt})   \\
\op{de}{V}{LR}         & (\bar d_{Lp} \gamma^\mu d_{Lr}) (\bar e_{Rs} \gamma_\mu e_{Rt})   \\
\op{\nu edu}{V}{LR}        & (\bar \nu_{Lp} \gamma^\mu e_{Lr})(\bar d_{Rs} \gamma_\mu u_{Rt})  +\hc \\
\op{uu}{V1}{LR}        & (\bar u_{Lp} \gamma^\mu u_{Lr})(\bar u_{Rs} \gamma_\mu u_{Rt})   \\
\op{uu}{V8}{LR}       & (\bar u_{Lp} \gamma^\mu T^A u_{Lr})(\bar u_{Rs} \gamma_\mu T^A u_{Rt})    \\ 
\op{ud}{V1}{LR}       & (\bar u_{Lp} \gamma^\mu u_{Lr}) (\bar d_{Rs} \gamma_\mu d_{Rt})  \\
\op{ud}{V8}{LR}       & (\bar u_{Lp} \gamma^\mu T^A u_{Lr})  (\bar d_{Rs} \gamma_\mu T^A d_{Rt})  \\
\op{du}{V1}{LR}       & (\bar d_{Lp} \gamma^\mu d_{Lr})(\bar u_{Rs} \gamma_\mu u_{Rt})   \\
\op{du}{V8}{LR}       & (\bar d_{Lp} \gamma^\mu T^A d_{Lr})(\bar u_{Rs} \gamma_\mu T^A u_{Rt}) \\
\op{dd}{V1}{LR}      & (\bar d_{Lp} \gamma^\mu d_{Lr})(\bar d_{Rs} \gamma_\mu d_{Rt})  \\
\op{dd}{V8}{LR}   & (\bar d_{Lp} \gamma^\mu T^A d_{Lr})(\bar d_{Rs} \gamma_\mu T^A d_{Rt}) \\
\op{uddu}{V1}{LR}   & (\bar u_{Lp} \gamma^\mu d_{Lr})(\bar d_{Rs} \gamma_\mu u_{Rt})  + \hc  \\
\op{uddu}{V8}{LR}      & (\bar u_{Lp} \gamma^\mu T^A d_{Lr})(\bar d_{Rs} \gamma_\mu T^A  u_{Rt})  + \hc \\
\end{array}
\end{align*}
\end{minipage}

\begin{minipage}[t]{3cm}
\renewcommand{\arraystretch}{1.51}
\small
\begin{align*}
\begin{array}[t]{c|c}
\multicolumn{2}{c}{\boldsymbol{(\overline L R)(\overline L R)+\hc}} \\
\hline
\op{ee}{S}{RR} 		& (\bar e_{Lp}   e_{Rr}) (\bar e_{Ls} e_{Rt})   \\
\op{eu}{S}{RR}  & (\bar e_{Lp}   e_{Rr}) (\bar u_{Ls} u_{Rt})   \\
\op{eu}{T}{RR} & (\bar e_{Lp}   \sigma^{\mu \nu}   e_{Rr}) (\bar u_{Ls}  \sigma_{\mu \nu}  u_{Rt})  \\
\op{ed}{S}{RR}  & (\bar e_{Lp} e_{Rr})(\bar d_{Ls} d_{Rt})  \\
\op{ed}{T}{RR} & (\bar e_{Lp} \sigma^{\mu \nu} e_{Rr}) (\bar d_{Ls} \sigma_{\mu \nu} d_{Rt})   \\
\op{\nu edu}{S}{RR} & (\bar   \nu_{Lp} e_{Rr})  (\bar d_{Ls} u_{Rt} ) \\
\op{\nu edu}{T}{RR} &  (\bar  \nu_{Lp}  \sigma^{\mu \nu} e_{Rr} )  (\bar  d_{Ls}  \sigma_{\mu \nu} u_{Rt} )   \\
\op{uu}{S1}{RR}  & (\bar u_{Lp}   u_{Rr}) (\bar u_{Ls} u_{Rt})  \\
\op{uu}{S8}{RR}   & (\bar u_{Lp}   T^A u_{Rr}) (\bar u_{Ls} T^A u_{Rt})  \\
\op{ud}{S1}{RR}   & (\bar u_{Lp} u_{Rr})  (\bar d_{Ls} d_{Rt})   \\
\op{ud}{S8}{RR}  & (\bar u_{Lp} T^A u_{Rr})  (\bar d_{Ls} T^A d_{Rt})  \\
\op{dd}{S1}{RR}   & (\bar d_{Lp} d_{Rr}) (\bar d_{Ls} d_{Rt}) \\
\op{dd}{S8}{RR}  & (\bar d_{Lp} T^A d_{Rr}) (\bar d_{Ls} T^A d_{Rt})  \\
\op{uddu}{S1}{RR} &  (\bar u_{Lp} d_{Rr}) (\bar d_{Ls}  u_{Rt})   \\
\op{uddu}{S8}{RR}  &  (\bar u_{Lp} T^A d_{Rr}) (\bar d_{Ls}  T^A u_{Rt})  \\[-0.5cm]
\end{array}
\end{align*}
\renewcommand{\arraystretch}{1.51}
\small
\begin{align*}
\begin{array}[t]{c|c}
\multicolumn{2}{c}{\boldsymbol{(\overline L R)(\overline R L) +\hc}} \\
\hline
\op{eu}{S}{RL}  & (\bar e_{Lp} e_{Rr}) (\bar u_{Rs}  u_{Lt})  \\
\op{ed}{S}{RL} & (\bar e_{Lp} e_{Rr}) (\bar d_{Rs} d_{Lt}) \\
\op{\nu edu}{S}{RL}  & (\bar \nu_{Lp} e_{Rr}) (\bar d_{Rs}  u_{Lt})  \\
\end{array}
\end{align*}
\end{minipage}
\end{adjustbox}
%%%
% --- END TABLE
%%%
\setlength{\abovecaptionskip}{0.15cm}
\setlength{\belowcaptionskip}{-3cm}
\label{tab:oplist1}
\centering
\caption{The non-four-fermions LEFT operators up to dimension 6 and the dimension-6 four-fermion LEFT operators conserving $B$ and $L$.}
\end{table}

\begin{table}[H]
%
% --- START TABLE
%%
\centering
\begin{minipage}[t]{3cm}
\renewcommand{\arraystretch}{1.5}
\small
\begin{align*}
\begin{array}[t]{c|c}
\multicolumn{2}{c}{\boldsymbol{\Delta L = 4 + \hc}}  \\
\hline
\op{\nu\nu}{S}{LL} &  (\nu_{Lp}^T C \nu_{Lr}^{}) (\nu_{Ls}^T C \nu_{Lt}^{} )  \\
\end{array}
\end{align*}
\end{minipage}
%%%
% --- END TABLE
%%%
%%

% --- START TABLE
%%
%\hspace{-1cm}
\begin{adjustbox}{width=\textwidth,center}
\begin{minipage}[t]{3cm}
\renewcommand{\arraystretch}{1.5}
\small
\begin{align*}
\begin{array}[t]{c|c}
\multicolumn{2}{c}{\boldsymbol{\Delta L =2 + \hc}}  \\
\hline
\op{\nu e}{S}{LL}  &  (\nu_{Lp}^T C \nu_{Lr}) (\bar e_{Rs} e_{Lt})   \\
\op{\nu e}{T}{LL} &  (\nu_{Lp}^T C \sigma^{\mu \nu} \nu_{Lr}) (\bar e_{Rs}\sigma_{\mu \nu} e_{Lt} )  \\
\op{\nu e}{S}{LR} &  (\nu_{Lp}^T C \nu_{Lr}) (\bar e_{Ls} e_{Rt} )  \\
\op{\nu u}{S}{LL}  &  (\nu_{Lp}^T C \nu_{Lr}) (\bar u_{Rs} u_{Lt} )  \\
\op{\nu u}{T}{LL}  &  (\nu_{Lp}^T C \sigma^{\mu \nu} \nu_{Lr}) (\bar u_{Rs} \sigma_{\mu \nu} u_{Lt} ) \\
\op{\nu u}{S}{LR}  &  (\nu_{Lp}^T C \nu_{Lr}) (\bar u_{Ls} u_{Rt} )  \\
\op{\nu d}{S}{LL}   &  (\nu_{Lp}^T C \nu_{Lr}) (\bar d_{Rs} d_{Lt} ) \\
\op{\nu d}{T}{LL}   &  (\nu_{Lp}^T C \sigma^{\mu \nu}  \nu_{Lr}) (\bar d_{Rs} \sigma_{\mu \nu} d_{Lt} ) \\
\op{\nu d}{S}{LR}  &  (\nu_{Lp}^T C \nu_{Lr}) (\bar d_{Ls} d_{Rt} ) \\
\op{\nu edu}{S}{LL} &  (\nu_{Lp}^T C e_{Lr}) (\bar d_{Rs} u_{Lt} )  \\
\op{\nu edu}{T}{LL}  & (\nu_{Lp}^T C  \sigma^{\mu \nu} e_{Lr}) (\bar d_{Rs}  \sigma_{\mu \nu} u_{Lt} ) \\
\op{\nu edu}{S}{LR}   & (\nu_{Lp}^T C e_{Lr}) (\bar d_{Ls} u_{Rt} ) \\
\op{\nu edu}{V}{RL}   & (\nu_{Lp}^T C \gamma^\mu e_{Rr}) (\bar d_{Ls} \gamma_\mu u_{Lt} )  \\
\op{\nu edu}{V}{RR}   & (\nu_{Lp}^T C \gamma^\mu e_{Rr}) (\bar d_{Rs} \gamma_\mu u_{Rt} )  \\
\end{array}
\end{align*}
\end{minipage}
%%%
% --- END TABLE
%%%
%
%%
%%
% --- START TABLE
%%
\begin{minipage}[t]{3cm}
\renewcommand{\arraystretch}{1.5}
\small
\begin{align*}
\begin{array}[t]{c|c}
\multicolumn{2}{c}{\boldsymbol{\Delta B = \Delta L = 1 + \hc}} \\
\hline
\op{udd}{S}{LL} &  \epsilon_{\alpha\beta\gamma}  (u_{Lp}^{\alpha T} C d_{Lr}^{\beta}) (d_{Ls}^{\gamma T} C \nu_{Lt}^{})   \\
\op{duu}{S}{LL} & \epsilon_{\alpha\beta\gamma}  (d_{Lp}^{\alpha T} C u_{Lr}^{\beta}) (u_{Ls}^{\gamma T} C e_{Lt}^{})  \\
\op{uud}{S}{LR} & \epsilon_{\alpha\beta\gamma}  (u_{Lp}^{\alpha T} C u_{Lr}^{\beta}) (d_{Rs}^{\gamma T} C e_{Rt}^{})  \\
\op{duu}{S}{LR} & \epsilon_{\alpha\beta\gamma}  (d_{Lp}^{\alpha T} C u_{Lr}^{\beta}) (u_{Rs}^{\gamma T} C e_{Rt}^{})   \\
\op{uud}{S}{RL} & \epsilon_{\alpha\beta\gamma}  (u_{Rp}^{\alpha T} C u_{Rr}^{\beta}) (d_{Ls}^{\gamma T} C e_{Lt}^{})   \\
\op{duu}{S}{RL} & \epsilon_{\alpha\beta\gamma}  (d_{Rp}^{\alpha T} C u_{Rr}^{\beta}) (u_{Ls}^{\gamma T} C e_{Lt}^{})   \\
\op{dud}{S}{RL} & \epsilon_{\alpha\beta\gamma}  (d_{Rp}^{\alpha T} C u_{Rr}^{\beta}) (d_{Ls}^{\gamma T} C \nu_{Lt}^{})   \\
\op{ddu}{S}{RL} & \epsilon_{\alpha\beta\gamma}  (d_{Rp}^{\alpha T} C d_{Rr}^{\beta}) (u_{Ls}^{\gamma T} C \nu_{Lt}^{})   \\
\op{duu}{S}{RR}  & \epsilon_{\alpha\beta\gamma}  (d_{Rp}^{\alpha T} C u_{Rr}^{\beta}) (u_{Rs}^{\gamma T} C e_{Rt}^{})  \\
\end{array}
\end{align*}
\end{minipage}
%%%
% --- END TABLE
%%%
%
%%
% --- START TABLE
%%
\begin{minipage}[t]{3cm}
\renewcommand{\arraystretch}{1.5}
\small
\begin{align*}
\begin{array}[t]{c|c}
\multicolumn{2}{c}{\boldsymbol{\Delta B = - \Delta L = 1 + \hc}}  \\
\hline
\op{ddd}{S}{LL} & \epsilon_{\alpha\beta\gamma}  (d_{Lp}^{\alpha T} C d_{Lr}^{\beta}) (\bar e_{Rs}^{} d_{Lt}^\gamma )  \\
\op{udd}{S}{LR}  & \epsilon_{\alpha\beta\gamma}  (u_{Lp}^{\alpha T} C d_{Lr}^{\beta}) (\bar \nu_{Ls}^{} d_{Rt}^\gamma )  \\
\op{ddu}{S}{LR} & \epsilon_{\alpha\beta\gamma}  (d_{Lp}^{\alpha T} C d_{Lr}^{\beta})  (\bar \nu_{Ls}^{} u_{Rt}^\gamma )  \\
\op{ddd}{S}{LR} & \epsilon_{\alpha\beta\gamma}  (d_{Lp}^{\alpha T} C d_{Lr}^{\beta}) (\bar e_{Ls}^{} d_{Rt}^\gamma ) \\
\op{ddd}{S}{RL}  & \epsilon_{\alpha\beta\gamma}  (d_{Rp}^{\alpha T} C d_{Rr}^{\beta}) (\bar e_{Rs}^{} d_{Lt}^\gamma )  \\
\op{udd}{S}{RR}  & \epsilon_{\alpha\beta\gamma}  (u_{Rp}^{\alpha T} C d_{Rr}^{\beta}) (\bar \nu_{Ls}^{} d_{Rt}^\gamma )  \\
\op{ddd}{S}{RR}  & \epsilon_{\alpha\beta\gamma}  (d_{Rp}^{\alpha T} C d_{Rr}^{\beta}) (\bar e_{Ls}^{} d_{Rt}^\gamma )  \\
\end{array}
\end{align*}
\end{minipage}
\end{adjustbox}
%%%
% --- END TABLE
%%%
%
\label{tab:oplist2}

\centering
\caption{The dimension-6 four-fermion LEFT operators violating $B$ and/or $L$.}
\end{table}

%\newpage
\section{SMEFT operators used in this paper}
\label{SMEFTops}

\subsection{Even-dimensional operators}
\label{evenSMEFTops}

These tables list the dimension-6 \cite{Grzadkowski:2010es} and dimension-8 \cite{Murphy:2020rsh} SMEFT operators that contribute to the matching conditions, separated into various categories.

\begin{table}[H]
\vspace{-0.75cm}
\begin{adjustbox}{width=\textwidth,center}

\begin{minipage}[t]{3cm}
\renewcommand{\arraystretch}{1.5}
\small
\begin{align*}
\begin{array}[t]{|c|c|}
\hline\multicolumn{2}{|c|}{\textbf{Classes $H^n$ and $H^nD^2$}} \\
\hline\textbf{Operator} & \textbf{WC} \\
\hline(H^\dagger H)^3 & \dfrac1{\Lambda^2}C_H \\[8pt]
\hline(H^\dagger H)^4 & \dfrac1{\Lambda^4}C_{H^8} \\[8pt]
\hline(H^\dagger H)\Box(H^\dagger H) & \dfrac1{\Lambda^2}C_{H\Box} \\[8pt]
\hline(H^\dagger D_\mu H)^*(H^\dagger D^\mu H) & \dfrac1{\Lambda^2}C_{HD} \\[8pt]
\hline(H^\dagger H)^2(D_\mu H^\dagger D^\mu H) & \dfrac1{\Lambda^4}C^{(1)}_{H^6} \\[8pt]
\hline(H^\dagger H)(H^\dagger\tau^IH)(D_\mu H^\dagger\tau^ID^\mu H) & \dfrac1{\Lambda^4}C^{(2)}_{H^6} \\[8pt]
\hline
\end{array}
\end{align*}
\end{minipage}

\begin{minipage}[t]{3cm}
\renewcommand{\arraystretch}{1.5}
\small
\begin{align*}
\begin{array}[t]{|c|c|}
\hline\multicolumn{2}{|c|}{\textbf{Classes $X^3H^n$}} \\
\hline\textbf{Operator} & \textbf{WC} \\
\hline f^{ABC}G_\mu^{A\nu}G_\nu^{B\rho}G_\rho^{C\mu} & \dfrac1{\Lambda^2}C_G \\[8pt]
\hline f^{ABC}\tilde G_\mu^{A\nu}G_\nu^{B\rho}G_\rho^{C\mu} & \dfrac1{\Lambda^2}C_{\tilde G} \\[8pt]
\hline f^{ABC}(H^\dagger H)G_\mu^{A\nu}G_\nu^{B\rho}G_\rho^{C\mu} & \dfrac1{\Lambda^2}C^{(1)}_{G^3H^2} \\[8pt]
\hline f^{ABC}(H^\dagger H)\tilde G_\mu^{A\nu}G_\nu^{B\rho}G_\rho^{C\mu} & \dfrac1{\Lambda^2}C^{(2)}_{G^3H^2} \\[8pt]
\hline
\end{array}
\end{align*}
\end{minipage}

\begin{minipage}[t]{3cm}
\renewcommand{\arraystretch}{1.5}
\small
\begin{align*}
\begin{array}[t]{|c|c|}
\hline\multicolumn{2}{|c|}{\textbf{Classes $\psi^2H^n$}} \\
\hline\textbf{Operator} & \textbf{WC} \\
\hline(H^\dagger H)(\overline l_pe_rH) & \dfrac1{\Lambda^2}C_{\substack{eH\\pr}} \\[8pt]
\hline(H^\dagger H)(\overline q_pu_rH) & \dfrac1{\Lambda^2}C_{\substack{uH\\pr}} \\[8pt]
\hline(H^\dagger H)(\overline q_pd_rH) & \dfrac1{\Lambda^2}C_{\substack{dH\\pr}} \\[8pt]
\hline(H^\dagger H)^2(\overline l_pe_rH) & \dfrac1{\Lambda^4}C_{\substack{leH^5\\pr}} \\[8pt]
\hline(H^\dagger H)^2(\overline q_pu_rH) & \dfrac1{\Lambda^4}C_{\substack{quH^5\\pr}} \\[8pt]
\hline(H^\dagger H)^2(\overline q_pd_rH) & \dfrac1{\Lambda^4}C_{\substack{qdH^5\\pr}} \\[8pt]
\hline
\end{array}
\end{align*}
\end{minipage}

\end{adjustbox}
\mbox{}\\[-0.75 cm]
\begin{adjustbox}{width=\textwidth,center}

\begin{minipage}[t]{3cm}
\renewcommand{\arraystretch}{1.5}
\small
\begin{align*}
\begin{array}[t]{|c|c|}
\hline\multicolumn{2}{|c|}{\textbf{Classes $X^2H^n$}} \\
\hline\textbf{Operator} & \textbf{WC} \\
\hline(H^\dagger H)(G^A_{\mu\nu}G^{A\mu\nu}) & \dfrac1{\Lambda^2}C_{HG} \\[8pt]
\hline(H^\dagger H)(W^I_{\mu\nu}W^{I\mu\nu}) & \dfrac1{\Lambda^2}C_{HW} \\[8pt]
\hline(H^\dagger H)(B_{\mu\nu}B^{\mu\nu}) & \dfrac1{\Lambda^2}C_{HB} \\[8pt]
\hline(H^\dagger\tau^IH)(W^I_{\mu\nu}B^{\mu\nu}) & \dfrac1{\Lambda^2}C_{HWB} \\[8pt]
\hline(H^\dagger H)^2G^A_{\mu\nu}G^{A\mu\nu} & \dfrac1{\Lambda^4}C^{(1)}_{G^2H^4} \\[8pt]
\hline(H^\dagger H)^2W^I_{\mu\nu}W^{I\mu\nu} & \dfrac1{\Lambda^4}C^{(1)}_{W^2H^4} \\[8pt]
\hline(H^\dagger\tau^IH)(H^\dagger\tau^JH)W^I_{\mu\nu}W^{J\mu\nu} & \dfrac1{\Lambda^4}C^{(3)}_{W^2H^4} \\[8pt]
\hline(H^\dagger H)(H^\dagger\tau^IH)W^I_{\mu\nu}B^{\mu\nu} & \dfrac1{\Lambda^4}C^{(1)}_{WBH^4} \\[8pt]
\hline(H^\dagger H)^2B_{\mu\nu}B^{\mu\nu} & \dfrac1{\Lambda^4}C^{(1)}_{B^2H^4} \\[8pt]
\hline
\end{array}
\end{align*}
\end{minipage}

\begin{minipage}[t]{3cm}
\renewcommand{\arraystretch}{1.5}
\small
\begin{align*}
\begin{array}[t]{|c|c|}
\hline\multicolumn{2}{|c|}{\textbf{Classes $\psi^2XH^n$}} \\
\hline\textbf{Operator} & \textbf{WC} \\
\hline(\overline l_p\sigma^{\mu\nu}e_r)\tau^IHW^I_{\mu\nu} & \dfrac1{\Lambda^2}C_{\substack{eW\\pr}} \\[8pt]
\hline(\overline l_p\sigma^{\mu\nu}e_r)HB_{\mu\nu} & \dfrac1{\Lambda^2}C_{\substack{eB\\pr}} \\[8pt]
\hline(\overline q_p\sigma^{\mu\nu}T^Au_r)\tilde HG^A_{\mu\nu} & \dfrac1{\Lambda^2}C_{\substack{uG\\pr}} \\[8pt]
\hline(\overline q_p\sigma^{\mu\nu}u_r)\tau^I\tilde HW^I_{\mu\nu} & \dfrac1{\Lambda^2}C_{\substack{uW\\pr}} \\[8pt]
\hline(\overline q_p\sigma^{\mu\nu}u_r)\tilde HB_{\mu\nu} & \dfrac1{\Lambda^2}C_{\substack{uB\\pr}} \\[8pt]
\hline(\overline q_p\sigma^{\mu\nu}T^Ad_r)HG^A_{\mu\nu} & \dfrac1{\Lambda^2}C_{\substack{dG\\pr}} \\
\hline
\end{array}
\end{align*}
\end{minipage}

\begin{minipage}[t]{3cm}
\renewcommand{\arraystretch}{1.5}
\small
\begin{align*}
\begin{array}[t]{|c|c|}
\hline\multicolumn{2}{|c|}{\textbf{Classes $\psi^2H^nD$}} \\
\hline\textbf{Operator} & \textbf{WC} \\
\hline(H^\dagger i\overleftrightarrow D_\mu H)(\overline l_p\gamma^\mu l_r) & \dfrac1{\Lambda^2}C^{(1)}_{\substack{Hl\\pr}} \\[8pt]
\hline(H^\dagger i\overleftrightarrow D^I_\mu H)(\overline l_p\tau^I\gamma^\mu l_r) & \dfrac1{\Lambda^2}C^{(3)}_{\substack{Hl\\pr}} \\[8pt]
\hline(H^\dagger i\overleftrightarrow D_\mu H)(\overline e_p\gamma^\mu e_r) & \dfrac1{\Lambda^2}C_{\substack{He\\pr}} \\[8pt]
\hline(H^\dagger i\overleftrightarrow D_\mu H)(\overline q_p\gamma^\mu q_r) & \dfrac1{\Lambda^2}C^{(1)}_{\substack{Hq\\pr}} \\[8pt]
\hline(H^\dagger i\overleftrightarrow D^I_\mu H)(\overline q_p\tau^I\gamma^\mu q_r) & \dfrac1{\Lambda^2}C^{(3)}_{\substack{Hq\\pr}} \\[8pt]
\hline(H^\dagger i\overleftrightarrow D_\mu H)(\overline u_p\gamma^\mu u_r) & \dfrac1{\Lambda^2}C_{\substack{Hu\\pr}} \\
\hline
\end{array}
\end{align*}
\end{minipage}

\end{adjustbox}
\centering
\caption{The even-dimensional non-four-fermion SMEFT operators appearing in this paper.}
\end{table}

\begin{table}[H]
\vspace{-0.75cm}
\begin{adjustbox}{width=\textwidth,center}

\begin{minipage}[t]{3cm}
\renewcommand{\arraystretch}{1.5}
\small
\begin{align*}
\begin{array}[t]{|c|c|}
\hline\multicolumn{2}{|c|}{\textbf{Classes $(\overline LL)(\overline LL)H^n$}} \\
\hline\textbf{Operator} & \textbf{WC} \\
\hline(\overline l_p\gamma_\mu l_r)(\overline l_s\gamma^\mu l_t) & \dfrac1{\Lambda^2}C_{\substack{ll\\prst}} \\[8pt]
\hline(\overline q_p\gamma_\mu q_r)(\overline q_s\gamma^\mu q_t) & \dfrac1{\Lambda^2}C^{(1)}_{\substack{qq\\prst}} \\[8pt]
\hline(\overline q_p\gamma_\mu\tau^Iq_r)(\overline q_s\gamma^\mu\tau^Iq_t) & \dfrac1{\Lambda^2}C^{(3)}_{\substack{qq\\prst}} \\[8pt]
\hline(\overline l_p\gamma_\mu l_r)(\overline q_s\gamma^\mu q_t) & \dfrac1{\Lambda^2}C^{(1)}_{\substack{lq\\prst}} \\[8pt]
\hline(\overline l_p\gamma_\mu\tau^Il_r)(\overline q_s\gamma^\mu\tau^Iq_t) & \dfrac1{\Lambda^2}C^{(3)}_{\substack{lq\\prst}} \\[8pt]
\hline(\overline l_p\gamma_\mu l_r)(\overline l_s\gamma^\mu l_t)(H^\dagger H) & \dfrac1{\Lambda^4}C^{(1)}_{\substack{l^4H^2\\prst}} \\
\hline
\end{array}
\end{align*}

\renewcommand{\arraystretch}{1.5}
\small
\begin{align*}
\begin{array}[t]{|c|c|}
\hline\multicolumn{2}{|c|}{\textbf{Classes $(\overline RR)(\overline RR)H^n$}} \\
\hline\textbf{Operator} & \textbf{WC} \\
\hline(\overline e_p\gamma_\mu e_r)(\overline e_s\gamma^\mu e_t) & \dfrac1{\Lambda^2}C_{\substack{ee\\prst}} \\[8pt]
\hline(\overline u_p\gamma_\mu u_r)(\overline u_s\gamma^\mu u_t) & \dfrac1{\Lambda^2}C_{\substack{uu\\prst}} \\[8pt]
\hline(\overline d_p\gamma_\mu d_r)(\overline d_s\gamma^\mu d_t) & \dfrac1{\Lambda^2}C_{\substack{dd\\prst}} \\[8pt]
\hline(\overline e_p\gamma_\mu e_r)(\overline u_s\gamma^\mu u_t) & \dfrac1{\Lambda^2}C_{\substack{eu\\prst}} \\[8pt]
\hline(\overline e_p\gamma_\mu e_r)(\overline d_s\gamma^\mu d_t) & \dfrac1{\Lambda^2}C_{\substack{ed\\prst}} \\[8pt]
\hline(\overline u_p\gamma_\mu u_r)(\overline d_s\gamma^\mu d_t) & \dfrac1{\Lambda^2}C^{(1)}_{\substack{ud\\prst}} \\[8pt]
\hline(\overline u_p\gamma_\mu T^Au_r)(\overline d_s\gamma^\mu T^Ad_t) & \dfrac1{\Lambda^2}C^{(8)}_{\substack{ud\\prst}} \\[8pt]
\hline(\overline e_p\gamma_\mu e_r)(\overline e_s\gamma^\mu e_t)(H^\dagger H) & \dfrac1{\Lambda^4}C_{\substack{e^4H^2\\prst}} \\[8pt]
\hline(\overline u_p\gamma_\mu u_r)(\overline u_s\gamma^\mu u_t)(H^\dagger H) & \dfrac1{\Lambda^4}C_{\substack{u^4H^2\\prst}} \\[8pt]
\hline(\overline d_p\gamma_\mu d_r)(\overline d_s\gamma^\mu d_t)(H^\dagger H) & \dfrac1{\Lambda^4}C_{\substack{d^4H^2\\prst}} \\[8pt]
\hline(\overline e_p\gamma_\mu e_r)(\overline u_s\gamma^\mu u_t)(H^\dagger H) & \dfrac1{\Lambda^4}C_{\substack{e^2u^2H^2\\prst}} \\[8pt]
\hline(\overline e_p\gamma_\mu e_r)(\overline d_s\gamma^\mu d_t)(H^\dagger H) & \dfrac1{\Lambda^4}C_{\substack{e^2d^2H^2\\prst}} \\[8pt]
\hline(\overline u_p\gamma_\mu u_r)(\overline d_s\gamma^\mu d_t)(H^\dagger H) & \dfrac1{\Lambda^4}C^{(1)}_{\substack{u^2d^2H^2\\prst}} \\[8pt]
\hline(\overline u_p\gamma_\mu T^Au_r)(\overline d_s\gamma^\mu T^Ad_t)(H^\dagger H) & \dfrac1{\Lambda^4}C^{(2)}_{\substack{u^2d^2H^2\\prst}} \\[8pt]
\hline
\end{array}
\end{align*}
\end{minipage}

\begin{minipage}[t]{3cm}
\renewcommand{\arraystretch}{1.5}
\small
\begin{align*}
\begin{array}[t]{|c|c|}
\hline\multicolumn{2}{|c|}{\textbf{Classes $(\overline LR)(\overline LR)H^n$}} \\
\hline\textbf{Operator} & \textbf{WC} \\
\hline(\overline q^j_pu_r)\epsilon_{jk}(\overline q^k_sd_t) & \dfrac1{\Lambda^2}C^{(1)}_{\substack{quqd\\prst}} \\[8pt]
\hline(\overline q^j_pT^Au_r)\epsilon_{jk}(\overline q^k_sT^Ad_t) & \dfrac1{\Lambda^2}C^{(8)}_{\substack{quqd\\prst}} \\[8pt]
\hline(\overline l^j_pe_r)\epsilon_{jk}(\overline q^k_su_t) & \dfrac1{\Lambda^2}C^{(1)}_{\substack{lequ\\prst}} \\[8pt]
\hline(\overline l^j_p\sigma_{\mu\nu}e_r)\epsilon_{jk}(\overline q^k_s\sigma^{\mu\nu}u_t) & \dfrac1{\Lambda^2}C^{(3)}_{\substack{lequ\\prst}} \\[8pt]
\hline(\overline q^j_pu_r)\epsilon_{jk}(\overline q^k_sd_t)(H^\dagger H) & \dfrac1{\Lambda^4}C^{(1)}_{\substack{q^2udH^2\\prst}} \\[8pt]
\hline(\overline q^j_pu_r)(\tau^I\epsilon)_{jk}(\overline q^k_sd_t)(H^\dagger\tau^IH) & \dfrac1{\Lambda^4}C^{(2)}_{\substack{q^2udH^2\\prst}} \\
\hline
\end{array}
\end{align*}

\renewcommand{\arraystretch}{1.5}
\small
\begin{align*}
\begin{array}[t]{|c|c|}
\hline\multicolumn{2}{|c|}{\textbf{Classes $(\slashed B)\psi^4H^n$}} \\
\hline\textbf{Operator} & \textbf{WC} \\
\hline\epsilon^{\alpha\beta\gamma}\epsilon^{jk}(d^{\alpha T}_pCu^\beta_r)(q^{\gamma T}_{js}Cl_{kt}) & \dfrac1{\Lambda^2}C_{\substack{duq\\prst}} \\[8pt]
\hline\epsilon^{\alpha\beta\gamma}\epsilon^{jk}(q^{\alpha T}_{jp}Cq^\beta_{kr})(u^{\gamma T}_sCe_t) & \dfrac1{\Lambda^2}C_{\substack{qqu\\prst}} \\[8pt]
\hline\epsilon^{\alpha\beta\gamma}\epsilon^{jn}\epsilon^{km}(q^{\alpha T}_{jp}Cq^\beta_{kr})(q^{\gamma T}_{ms}Cl_{nt}) & \dfrac1{\Lambda^2}C_{\substack{qqq\\prst}} \\[8pt]
\hline\epsilon^{\alpha\beta\gamma}(d^{\alpha T}_pCu^\beta_r)(u^{\gamma T}_sCe_t) & \dfrac1{\Lambda^2}C_{\substack{duu\\prst}} \\[8pt]
\hline\epsilon^{\alpha\beta\gamma}\epsilon^{jk}(d^{\alpha T}_pCu^\beta_r)(q^{\gamma T}_{js}Cl_{kt})(H^\dagger H) & \dfrac1{\Lambda^4}C^{(1)}_{\substack{lqudH^2\\prst}} \\[8pt]
\hline\epsilon^{\alpha\beta\gamma}(\epsilon\tau^I)^{jk}(d^{\alpha T}_pCu^\beta_r)(q^{\gamma T}_{js}Cl_{kt})(H^\dagger\tau^IH) & \dfrac1{\Lambda^4}C^{(2)}_{\substack{lqudH^2\\prst}} \\
\hline
\end{array}
\end{align*}
\end{minipage}

\begin{minipage}[t]{3cm}
\renewcommand{\arraystretch}{1.5}
\small
\begin{align*}
\begin{array}[t]{|c|c|}
\hline\multicolumn{2}{|c|}{\textbf{Classes $(\overline LL)(\overline RR)H^n$}} \\
\hline\textbf{Operator} & \textbf{WC} \\
\hline(\overline l_p\gamma_\mu l_r)(\overline e_s\gamma^\mu e_t) & \dfrac1{\Lambda^2}C_{\substack{le\\prst}} \\[8pt]
\hline(\overline l_p\gamma_\mu l_r)(\overline u_s\gamma^\mu u_t) & \dfrac1{\Lambda^2}C_{\substack{lu\\prst}} \\[8pt]
\hline(\overline l_p\gamma_\mu l_r)(\overline d_s\gamma^\mu d_t) & \dfrac1{\Lambda^2}C_{\substack{ld\\prst}} \\[8pt]
\hline(\overline q_p\gamma_\mu q_r)(\overline e_s\gamma^\mu e_t) & \dfrac1{\Lambda^2}C_{\substack{qe\\prst}} \\[8pt]
\hline(\overline q_p\gamma_\mu q_r)(\overline u_s\gamma^\mu u_t) & \dfrac1{\Lambda^2}C^{(1)}_{\substack{qu\\prst}} \\[8pt]
\hline(\overline q_p\gamma_\mu T^Aq_r)(\overline u_s\gamma^\mu T^Au_t) & \dfrac1{\Lambda^2}C^{(8)}_{\substack{qu\\prst}} \\
\hline
\end{array}
\end{align*}

\renewcommand{\arraystretch}{1.5}
\small
\begin{align*}
\begin{array}[t]{|c|c|}
\hline\multicolumn{2}{|c|}{\textbf{Classes $(\overline LR)(\overline RL)H^n$}} \\
\hline\textbf{Operator} & \textbf{WC} \\
\hline(\overline l^j_pe_r)(\overline d_sq^j_t) & \dfrac1{\Lambda^2}C_{\substack{ledq\\prst}} \\[8pt]
\hline(\overline l^j_pe_r)(\overline d_sq^j_t)(H^\dagger H) & \dfrac1{\Lambda^4}C^{(1)}_{\substack{leqdH^2\\prst}} \\[8pt]
\hline(\overline l_pe_r)\tau^I(\overline d_sq_t)(H^\dagger\tau^IH) & \dfrac1{\Lambda^4}C^{(2)}_{\substack{leqdH^2\\prst}} \\[8pt]
\hline(\overline l_pd_rH)(\tilde H^\dagger\overline u_sl_t) & \dfrac1{\Lambda^4}C_{\substack{l^2udH^2\\prst}} \\[8pt]
\hline(\overline l_pe_rH)(\tilde H^\dagger\overline u_sq_t) & \dfrac1{\Lambda^4}C^{(5)}_{\substack{lequH^2\\prst}} \\[8pt]
\hline(\overline q_pd_rH)(\tilde H^\dagger\overline u_sq_t) & \dfrac1{\Lambda^4}C^{(5)}_{\substack{q^2udH^2\\prst}} \\[8pt]
\hline(\overline q_pT^Ad_rH)(\tilde H^\dagger\overline u_sT^Aq_t) & \dfrac1{\Lambda^4}C^{(6)}_{\substack{q^2udH^2\\prst}} \\[8pt]
\hline
\end{array}
\end{align*}
\end{minipage}

\end{adjustbox}
\centering
\caption{The even-dimensional four-fermion SMEFT operators appearing in this paper.}
\end{table}

\subsection{Odd-dimensional operators}

There is only one dimension-5 SMEFT operator: $\epsilon^{ij}\epsilon^{kl}(l^T_{ip}Cl_{kr})H_jH_l$. The basis for the dimension-7 operators used here is equivalent to those given in Refs.~\cite{Lehman, XiaoDongMa}.

\begin{table}[H]
\vspace{-0.75cm}
\begin{adjustbox}{width=\textwidth,center}

\begin{minipage}[t]{3cm}
\renewcommand{\arraystretch}{1.5}
\small
\begin{align*}
\begin{array}[t]{|c|c|}
\hline\multicolumn{2}{|c|}{\textbf{Classes $\psi^2H^n$}} \\
\hline\textbf{Operator} & \textbf{WC} \\
\hline\epsilon^{ij}\epsilon^{kl}(l^T_{ip}Cl_{kr})H_jH_l & \dfrac1{\Lambda}C_{\substack{5\\pr}} \\[8pt]
\hline\epsilon^{ij}\epsilon^{kl}(l^T_{ip}Cl_{kr})H_jH_l(H^\dagger H) & \dfrac1{\Lambda^3}C_{\substack{l^2H^4\\pr}} \\[8pt]
\hline
\end{array}
\end{align*}

\renewcommand{\arraystretch}{1.5}
\small
\begin{align*}
\begin{array}[t]{|c|c|}
\hline\multicolumn{2}{|c|}{\textbf{Class $\psi^2H^3D$}} \\
\hline\textbf{Operator} & \textbf{WC} \\
\hline i\epsilon^{ij}\epsilon^{kl}(l^T_{ip}C\gamma^\mu e_r)H_jH_k(D_\mu H)_l & \dfrac1{\Lambda^3}C_{\substack{leH^3D\\pr}} \\[8pt]
\hline
\end{array}
\end{align*}

\renewcommand{\arraystretch}{1.5}
\small
\begin{align*}
\begin{array}[t]{|c|c|}
\hline\multicolumn{2}{|c|}{\textbf{Class $\psi^2H^2X$}} \\
\hline\textbf{Operator} & \textbf{WC} \\
\hline\epsilon^{ij}\epsilon^{kl}(l^T_{ip}C\sigma_{\mu\nu}l_{kr})H_jH_lB^{\mu\nu} & \dfrac1{\Lambda^3}C_{\substack{l^2H^2B\\pr}} \\[8pt]
\hline\epsilon^{ij}(\epsilon\tau^I)^{kl}(l^T_{ip}C\sigma_{\mu\nu}l_{kr})H_jH_lW^{I\mu\nu} & \dfrac1{\Lambda^3}C_{\substack{l^2H^2W\\pr}} \\[8pt]
\hline
\end{array}
\end{align*}
\end{minipage}

\begin{minipage}[t]{3cm}
\renewcommand{\arraystretch}{1.5}
\small
\begin{align*}
\begin{array}[t]{|c|c|}
\hline\multicolumn{2}{|c|}{\textbf{Class $\psi^4H$}} \\
\hline\textbf{Operator} & \textbf{WC} \\
\hline\epsilon^{ij}\epsilon^{kl}(l^T_{ip}Cl_{kr})(\overline e_sl_{jt})H_l & \dfrac1{\Lambda^3}C_{\substack{l^3eH\\prst}} \\[8pt]
\hline\epsilon^{ij}\epsilon^{kl}(l^T_{ip}Cl_{kr})(\overline d_sq_{lt})H_j & \dfrac1{\Lambda^3}C^{(1)}_{\substack{l^2dqH\\prst}} \\[8pt]
\hline\epsilon^{ij}\epsilon^{kl}(l^T_{ip}C\sigma_{\mu\nu}l_{kr})(\overline d_s\sigma^{\mu\nu}q_{lt})H_j & \dfrac1{\Lambda^3}C^{(2)}_{\substack{l^2dqH\\prst}} \\[8pt]
\hline\epsilon^{ij}(l^T_{ip}Cl_{kr})(\overline q^k_su_t)H_j & \dfrac1{\Lambda^3}C_{\substack{l^2quH\\prst}} \\[8pt]
\hline\epsilon^{\alpha\beta\gamma}\epsilon^{ij}(q^{\alpha T}_{kp}Cq^\beta_{ir})(\overline l^k_sd^\gamma_t)\tilde H_j & \dfrac1{\Lambda^3}C_{\substack{q^2ldH\\prst}} \\[8pt]
\hline\epsilon^{\alpha\beta\gamma}(d^{\alpha T}_pCd^\beta_r)(\overline l_sd^\gamma_t)H & \dfrac1{\Lambda^3}C_{\substack{d^3lH\\prst}} \\[8pt]
\hline\epsilon^{\alpha\beta\gamma}(u^{\alpha T}_pCd^\beta_r)(\overline l_sd^\gamma_t)\tilde H & \dfrac1{\Lambda^3}C_{\substack{ud^2lH\\prst}} \\[8pt]
\hline\epsilon^{ij}(l^T_{ip}C\gamma_\mu e_r)(\overline d_s\gamma^\mu u_t)H_j & \dfrac1{\Lambda^3}C_{\substack{leduH\\prst}} \\[8pt]
\hline\epsilon^{\alpha\beta\gamma}\epsilon^{ij}(d^{\alpha T}_pCd^\beta_r)(\overline e_sq^\gamma_{it})\tilde H_j & \dfrac1{\Lambda^3}C_{\substack{eqd^2H\\prst}} \\[8pt]
\hline
\end{array}
\end{align*}
\end{minipage}

\end{adjustbox}
\centering
\caption{The odd-dimensional SMEFT operators appearing in this paper.}
\end{table}

\section{Useful Fierz identities}
\label{Fierz}

The following Fierz identities are needed to derive the matching conditions given in this paper.
\subsection{For $(\overline LL)(\overline LL)$ and $(\overline RR)(\overline RR)$ operators}
In the case of a four-lepton operator, the identities take the form
\beq
(\overline\nu_{Lp}\gamma^\mu e_{Lt})(\overline e_{Ls}\gamma_\mu\nu_{Lr}) = (\overline\nu_{Lp}\gamma^\mu\nu_{Lr})(\overline e_{Ls}\gamma_\mu e_{Lt}).
\eeq
In the case of a four-quark operator, color has to be taken into consideration. This is done through the identity $\delta_{\alpha\lambda}\delta_{\kappa\beta}=2T^A_{\alpha\beta}T^A_{\kappa\lambda}+\dfrac13\delta_{\alpha\beta}\delta_{\kappa\lambda}$. The identities are (for instance)
\beq
(\overline u_{Lp}\gamma^\mu d_{Lt})(\overline d_{Ls}\gamma_\mu u_{Lr}) = 2(\overline u_{Lp}\gamma^\mu T^Au_{Lr})(\overline d_{Ls}\gamma_\mu T^Ad_{Lt}) + \dfrac13(\overline u_{Lp}\gamma^\mu u_{Lr})(\overline d_{Ls}\gamma_\mu d_{Lt}).
\eeq
\subsection{For $(\overline LL)(\overline RR)$ operators}
In the case of a four-lepton operator or a two-lepton and two-quark operator, the Fierz identities take the form
\beq
(\overline\nu_{Lp}e_{Rt})(\overline e_{Rs}\nu_{Lr})=-\dfrac12(\overline\nu_{Lp}\gamma^\mu\nu_{Lr})(\overline e_{Rs}\gamma_\mu e_{Rt}) ~.
\eeq
In the case of a four-quark operator, the identities take the following forms:
\beq
(\overline u_{Lp}d_{Rt})(\overline d_{Rs}u_{Lr}) = -(\overline u_{Lp}\gamma^\mu T^Au_{Lr})(\overline d_{Rs}\gamma_\mu T^Ad_{Rt})-\dfrac16(\overline u_{Lp}\gamma^\mu u_{Lr})(\overline d_{Rs}\gamma_\mu d_{Rt}) ~,
\eeq
\beq
(\overline u_{Lp}T^Ad_{Rt})(\overline d_{Rs}T^Au_{Lr}) = -\dfrac29(\overline u_{Lp}\gamma^\mu u_{Lr})(\overline d_{Rs}\gamma_\mu d_{Rt})+\dfrac16(\overline u_{Lp}\gamma^\mu T^Au_{Lr})(\overline d_{Rs}\gamma_\mu T^Ad_{Rt}) ~.
\eeq
\subsection{For fermion-number-violating operators}
The needed identities take the form
\beq
(\nu^T_{Lr}C\nu_{Lr})(\overline e_{Rs}e_{Lt})=-\dfrac12(\nu^T_{Lp}Ce_{Lt})(\overline e_{Rs}\nu_{Lr})-\dfrac18(\nu^T_{Lp}C\sigma_{\mu\nu}e_{Lt})(\overline e_{Rs}\sigma^{\mu\nu}\nu_{Lr}) ~,
\eeq
\beq
(\nu^T_{Lr}C\nu_{Lr})(\overline e_{Ls}e_{Rt})=-\dfrac12(\nu^T_{Lp}C\gamma_\mu e_{Rt})(\overline e_{Rs}\gamma^\mu\nu_{Rr}) ~,
\eeq
\beq
(\nu^T_{Lr}C\nu_{Lr})(e^T_{Ls}Ce_{Lt})=-\dfrac12(\nu^T_{Lp}Ce_{Lt})(e^T_{Ls}C\nu_{Lr})-\dfrac18(\nu^T_{Lp}C\sigma_{\mu\nu}e_{Lt})(e^T_{Ls}C\sigma^{\mu\nu}\nu_{Lr}) ~.
\eeq

\section{Matching conditions}
\label{Matchings}

\subsection{$\nu\nu+\,$h.c. operator}
\begin{center}
\begin{tabular}{|c|c|}
\hline\textbf{LEFT WC} & \textbf{Matching} \\
\hline$\Lambda C_{\substack{\nu\\pr}}$ & $\dfrac{{v_T}^2}{2\Lambda}\left[C_{\substack{5\\pr}}+\dfrac{{v_T}^2}{2\Lambda^2}C_{\substack{l^2H^4\\pr}}\right]$ \\[8pt]
\hline
\end{tabular}
\end{center}
\subsection{$(\nu\nu)X+\,$h.c. and $(\overline LR)X+\,$h.c. operators}
\begin{center}
\begin{tabular}{|c|c|}
\hline\textbf{LEFT WC ($+\,$c.c.)} & \textbf{Matching ($+\,$c.c.)} \\
\hline$\dfrac1{\Lambda^2}C_{\substack{\nu\gamma\\pr}}$ & $\dfrac{{v_T}^2}{2g_Z\Lambda^3}\left[gC_{\substack{l^2H^2B\\pr}}-\dfrac{g'}2\left(C_{\substack{l^2H^2W\\pr}}-C_{\substack{l^2H^2W\\rp}}\right)\right]$ \\[8pt]
\hline$\dfrac1{\Lambda^2}C_{\substack{e\gamma\\pr}}$ & $\dfrac{v_T}{\sqrt2\,g_Z\,\Lambda^2}\left[\begin{aligned}
& \left(gC_{\substack{eB\\pr}}-g'C_{\substack{eW\\pr}}\right) \\
& +\dfrac{{v_T}^2}{2\Lambda^2}\left(gC_{\substack{leBH^3\\pr}}-g'C_{\substack{leWH^3\\pr}}^{(1)}-g'C_{\substack{leWH^3\\pr}}^{(2)}\right)
\end{aligned}\right]$ \\[24pt]
\hline$\dfrac1{\Lambda^2}C_{\substack{u\gamma\\pr}}$ & $\dfrac{v_T}{\sqrt2\,g_Z\,\Lambda^2}\left[\begin{aligned}
& \left(gC_{\substack{uB\\pr}}+g'C_{\substack{uW\\pr}}\right) \\
& +\dfrac{{v_T}^2}{2\Lambda^2}\left(gC_{\substack{quBH^3\\pr}}+g'C_{\substack{quWH^3\\pr}}^{(1)}-g'C_{\substack{quWH^3\\pr}}^{(2)}\right)
\end{aligned}\right]$ \\[24pt]
\hline$\dfrac1{\Lambda^2}C_{\substack{d\gamma\\pr}}$ & $\dfrac{v_T}{\sqrt2\,g_Z\,\Lambda^2}\left[\begin{aligned}
& \left(gC_{\substack{dB\\pr}}-g'C_{\substack{dW\\pr}}\right) \\
& +\dfrac{{v_T}^2}{2\Lambda^2}\left(gC_{\substack{qdBH^3\\pr}}-g'C_{\substack{qdWH^3\\pr}}^{(1)}-g'C_{\substack{qdWH^3\\pr}}^{(2)}\right)
\end{aligned}\right]$ \\[24pt]
\hline$\dfrac1{\Lambda^2}C_{\substack{uG\\pr}}$ & $\dfrac{v_T}{\sqrt2\,\Lambda^2}\left[C_{\substack{uG\\pr}}+\dfrac{{v_T}^2}{2\Lambda^2}C_{\substack{quGH^3\\pr}}\right]$ \\[8pt]
\hline$\dfrac1{\Lambda^2}C_{\substack{dG\\pr}}$ & $\dfrac{v_T}{\sqrt2\,\Lambda^2}\left[C_{\substack{dG\\pr}}+\dfrac{{v_T}^2}{2\Lambda^2}C_{\substack{qdGH^3\\pr}}\right]$ \\[8pt]
\hline
\end{tabular}
\end{center}
The non-physical ratios $g/g_Z$ and $g'/g_Z$ appearing here can be expressed in terms of the corrected coupling constants $\overline g$ and $\overline g'$ and the SMEFT WC's, using the following equations:
\beq
\dfrac{g}{g_Z}=\dfrac{\overline g}{\sqrt{\overline g^2+\overline g'^2}}\left[\begin{aligned}
& 1+\dfrac{\overline g'^2{v_T}^2}{(\overline g^2+\overline g'^2)\Lambda^2}(C_{HB}-C_{HW}) \\
& +\dfrac{\overline g'^2{v_T}^4}{2(\overline g^2+\overline g'^2)\Lambda^4}(C_{B^2H^4}^{(1)}-C_{W^2H^4}^{(1)}) \\
& +\dfrac{{v_T}^4}{2(\overline g^2+\overline g'^2)^2\Lambda^4}\left(\begin{aligned}
& 3\overline g'^4\,[C_{HB}]^2-\overline g'^2[4\overline g^2+\overline g'^2]\,[C_{HW}]^2 \\
& +2\overline g'^2[2\overline g^2-\overline g'^2]\,C_{HW}\,C_{HB}
\end{aligned}\right)
\end{aligned}\right] ~,
\eeq
\beq
\dfrac{g'}{g_Z}=\dfrac{\overline g'}{\sqrt{\overline g^2+\overline g'^2}}\left[\begin{aligned}
& 1+\dfrac{\overline g^2{v_T}^2}{(\overline g^2+\overline g'^2)\Lambda^2}(C_{HW}-C_{HB}) \\
& +\dfrac{\overline g^2{v_T}^4}{2(\overline g^2+\overline g'^2)\Lambda^4}(C_{W^2H^4}^{(1)}-C_{B^2H^4}^{(1)}) \\
& +\dfrac{{v_T}^4}{2(\overline g^2+\overline g'^2)^2\Lambda^4}\left(\begin{aligned}
& 3\overline g^4\,[C_{HW}]^2-\overline g^2[4\overline g'^2+\overline g^2]\,[C_{HB}]^2 \\
& +2\overline g^2[2\overline g'^2-\overline g^2]\,C_{HW}\,C_{HB}
\end{aligned}\right)
\end{aligned}\right] ~.
\eeq
\subsection{$X^3$ operators}
\begin{center}
\begin{tabular}{|c|c|}
\hline\textbf{LEFT WC} & \textbf{Matching} \\
\hline$\dfrac1{\Lambda^2}\mathcal C_G$ & $\dfrac1{\Lambda^2}\left[C_G+\dfrac{{v_T}^2}{2\Lambda^2}C^{(1)}_{G^3H^2}\right]$ \\[8pt]
\hline$\dfrac1{\Lambda^2}\mathcal C_{\tilde G}$ & $\dfrac1{\Lambda^2}\left[C_{\tilde G}+\dfrac{{v_T}^2}{2\Lambda^2}C^{(2)}_{G^3H^2}\right]$ \\[8pt]
\hline
\end{tabular}
\end{center}
\subsection{$(\overline LL)(\overline LL)$ operators}
\begin{center}
\begin{tabular}{|c|c|}
\hline\textbf{LEFT WC} & \textbf{Matching} \\
\hline$\dfrac1{\Lambda^2}C_{\substack{\nu\nu\\prst}}^{V,LL}$ & $\dfrac1{\Lambda^2}\left[C_{\substack{ll\\prst}}+\dfrac{{v_T}^2}{2\Lambda^2}\left(C_{\substack{l^4H^2\\prst}}^{(1)}-C_{\substack{l^4H^2\\prst}}^{(2)}-C_{\substack{l^4H^2\\stpr}}^{(2)}\right)\right]$ \\[12pt]
& $-\dfrac{\overline{g_Z}^2}{4{M_Z}^2}\left([Z_{\nu_L}]^{\text{eff}}_{pr}[Z_{\nu_L}]^{\text{eff}}_{st}+[Z_{\nu_L}]^{\text{eff}}_{pt}[Z_{\nu_L}]^{\text{eff}}_{sr}\right)$ \\[8pt]
\hline$\dfrac1{\Lambda^2}C_{\substack{ee\\prst}}^{V,LL}$ & $\dfrac1{\Lambda^2}\left[C_{\substack{ll\\prst}}+\dfrac{{v_T}^2}{2\Lambda^2}\left(C_{\substack{l^4H^2\\prst}}^{(1)}+C_{\substack{l^4H^2\\prst}}^{(2)}+C_{\substack{l^4H^2\\stpr}}^{(2)}\right)\right]$ \\[12pt]
& $-\dfrac{\overline{g_Z}^2}{4{M_Z}^2}\left([Z_{e_L}]^{\text{eff}}_{pr}[Z_{e_L}]^{\text{eff}}_{st}+[Z_{e_L}]^{\text{eff}}_{pt}[Z_{e_L}]^{\text{eff}}_{sr}\right)$ \\[8pt]
\hline$\dfrac1{\Lambda^2}C_{\substack{\nu e\\prst}}^{V,LL}$ & $\dfrac1{\Lambda^2}\left[\left(C_{\substack{ll\\prst}}+C_{\substack{ll\\stpr}}\right)+\dfrac{{v_T}^2}{2\Lambda^2}\left(C_{\substack{l^4H^2\\prst}}^{(1)}+C_{\substack{l^4H^2\\stpr}}^{(1)}+C_{\substack{l^4H^2\\prst}}^{(2)}-C_{\substack{l^4H^2\\stpr}}^{(2)}\right)\right]$ \\[12pt]
& $-\dfrac{\overline{g_Z}^2}{{M_Z}^2}[Z_{\nu_L}]^{\text{eff}}_{pr}[Z_{e_L}]^{\text{eff}}_{st}-\dfrac{\overline g^2}{2{M_W}^2}[W_l]^{\text{eff}}_{pt}{[W_l]^{\text{eff}}_{rs}}^*$ \\[8pt]
\hline$\dfrac1{\Lambda^2}C_{\substack{uu\\prst}}^{V,LL}$ & $\dfrac1{\Lambda^2}\left[\left(C^{(1)}_{\substack{qq\\prst}}+C^{(3)}_{\substack{qq\\prst}}\right)+\dfrac{{v_T}^2}{2\Lambda^2}\left(C_{\substack{q^4H^2\\prst}}^{(1)}-C_{\substack{q^4H^2\\prst}}^{(2)}-C_{\substack{q^4H^2\\stpr}}^{(2)}+C_{\substack{q^4H^2\\prst}}^{(3)}\right)\right]$ \\
\hline
\end{tabular}
\end{center}
\begin{center}
\begin{tabular}{|c|c|}
\hline\textbf{LEFT WC} & \textbf{Matching} \\
\hline$\dfrac1{\Lambda^2}C_{\substack{ud\\prst}}^{V1,LL}$ & $\dfrac1{\Lambda^2}\left[\begin{aligned}
& \left(C^{(1)}_{\substack{qq\\prst}}+C^{(1)}_{\substack{qq\\stpr}}-C^{(3)}_{\substack{qq\\prst}}-C^{(3)}_{\substack{qq\\stpr}}+\dfrac23C^{(3)}_{\substack{qq\\ptsr}}+\dfrac23C^{(3)}_{\substack{qq\\srpt}}\right) \\
& +\dfrac{{v_T}^2}{2\Lambda^2}\left(\begin{aligned}
& C_{\substack{q^4H^2\\prst}}^{(1)}+C_{\substack{q^4H^2\\stpr}}^{(1)}+C_{\substack{q^4H^2\\prst}}^{(2)}-C_{\substack{q^4H^2\\stpr}}^{(2)}-C_{\substack{q^4H^2\\prst}}^{(3)}-C_{\substack{q^4H^2\\stpr}}^{(3)} \\
& +\dfrac23C_{\substack{q^4H^2\\ptsr}}^{(3)}+\dfrac23C_{\substack{q^4H^2\\srpt}}^{(3)}+\dfrac{2i}3C_{\substack{q^4H^2\\ptsr}}^{(5)}-\dfrac{2i}3C_{\substack{q^4H^2\\srpt}}^{(5)}
\end{aligned}\right)
\end{aligned}\right]$ \\[12pt]
& $-\dfrac{\overline{g_Z}^2}{{M_Z}^2}[Z_{u_L}]^{\text{eff}}_{pr}[Z_{d_L}]^{\text{eff}}_{st}-\dfrac{\overline g^2}{6{M_W}^2}[W_q]^{\text{eff}}_{pt}{[W_q]^{\text{eff}}_{rs}}^*$ \\[8pt]
\hline$\dfrac1{\Lambda^2}C_{\substack{ud\\prst}}^{V8,LL}$ & $\dfrac4{\Lambda^2}\left[\left(C^{(3)}_{\substack{qq\\ptsr}}+C^{(3)}_{\substack{qq\\srpt}}\right)+\dfrac{{v_T}^2}{2\Lambda^2}\left(C_{\substack{q^4H^2\\ptsr}}^{(3)}+C_{\substack{q^4H^2\\srpt}}^{(3)}-iC_{\substack{q^4H^2\\prst}}^{(5)}+iC_{\substack{q^4H^2\\stpr}}^{(5)}\right)\right]$ \\[12pt]
& $-\dfrac{\overline g^2}{{M_W}^2}[W_q]^{\text{eff}}_{pt}{[W_q]^{\text{eff}}_{rs}}^*$ \\[8pt]
\hline$\dfrac1{\Lambda^2}C_{\substack{\nu u\\prst}}^{V,LL}$ & $\dfrac1{\Lambda^2}\left[\left(C^{(1)}_{\substack{lq\\prst}}+C^{(3)}_{\substack{lq\\prst}}\right)+\dfrac{{v_T}^2}{2\Lambda^2}\left(C_{\substack{l^2q^2H^2\\prst}}^{(1)}-C_{\substack{l^2q^2H^2\\prst}}^{(2)}+C_{\substack{l^2q^2H^2\\prst}}^{(3)}-C_{\substack{l^2q^2H^2\\prst}}^{(4)}\right)\right]$ \\[12pt]
& $-\dfrac{\overline{g_Z}^2}{{M_Z}^2}\,[Z_{\nu_L}]^{\text{eff}}_{pr}[Z_{u_L}]^{\text{eff}}_{st}$ \\
\hline
\end{tabular}
\end{center}
\subsection{$(\overline RR)(\overline RR)$ operators}
\begin{center}
\begin{tabular}{|c|c|}
\hline\textbf{LEFT WC} & \textbf{Matching} \\
\hline$\dfrac1{\Lambda^2}C_{\substack{ee\\prst}}^{V,RR}$ & $\dfrac1{\Lambda^2}\left[C_{\substack{ee\\prst}}+\dfrac{{v_T}^2}{2\Lambda^2}C_{\substack{e^4H^2\\prst}}\right]-\dfrac{\overline{g_Z}^2}{4{M_Z}^2}\left([Z_{e_R}]^{\text{eff}}_{pr}[Z_{e_R}]^{\text{eff}}_{st}+[Z_{e_R}]^{\text{eff}}_{pt}[Z_{e_R}]^{\text{eff}}_{sr}\right)$ \\[8pt]
\hline$\dfrac1{\Lambda^2}C_{\substack{eu\\prst}}^{V,RR}$ & $\dfrac1{\Lambda^2}\left[C_{\substack{eu\\prst}}+\dfrac{{v_T}^2}{2\Lambda^2}C_{\substack{e^2u^2H^2\\prst}}\right]-\dfrac{\overline{g_Z}^2}{{M_Z}^2}[Z_{e_R}]^{\text{eff}}_{pr}[Z_{u_R}]^{\text{eff}}_{st}$ \\[8pt]
\hline$\dfrac1{\Lambda^2}C_{\substack{ed\\prst}}^{V,RR}$ & $\dfrac1{\Lambda^2}\left[C_{\substack{ed\\prst}}+\dfrac{{v_T}^2}{2\Lambda^2}C_{\substack{e^2d^2H^2\\prst}}\right]-\dfrac{\overline{g_Z}^2}{{M_Z}^2}[Z_{e_R}]^{\text{eff}}_{pr}[Z_{d_R}]^{\text{eff}}_{st}$ \\[8pt]
\hline$\dfrac1{\Lambda^2}C_{\substack{uu\\prst}}^{V,RR}$ & $\dfrac1{\Lambda^2}\left[C_{\substack{uu\\prst}}+\dfrac{{v_T}^2}{2\Lambda^2}C_{\substack{u^4H^2\\prst}}\right]-\dfrac{\overline{g_Z}^2}{2{M_Z}^2}\,[Z_{u_R}]^{\text{eff}}_{pr}[Z_{u_R}]^{\text{eff}}_{st}$ \\[8pt]
\hline$\dfrac1{\Lambda^2}C_{\substack{dd\\prst}}^{V,RR}$ & $\dfrac1{\Lambda^2}\left[C_{\substack{dd\\prst}}+\dfrac{{v_T}^2}{2\Lambda^2}C_{\substack{d^4H^2\\prst}}\right]-\dfrac{\overline{g_Z}^2}{2{M_Z}^2}\,[Z_{d_R}]^{\text{eff}}_{pr}[Z_{d_R}]^{\text{eff}}_{st}$ \\[8pt]
\hline$\dfrac1{\Lambda^2}C_{\substack{ud\\prst}}^{V1,RR}$ & $\dfrac1{\Lambda^2}\left[C^{(1)}_{\substack{ud\\prst}}+\dfrac{{v_T}^2}{2\Lambda^2}C_{\substack{u^2d^2H^2\\prst}}^{(1)}\right]-\dfrac{\overline{g_Z}^2}{{M_Z}^2}[Z_{u_R}]^{\text{eff}}_{pr}[Z_{d_R}]^{\text{eff}}_{st}-\dfrac{\overline g^2}{6{M_W}^2}[W_R]^{\text{eff}}_{pt}{[W_R]^{\text{eff}}_{rs}}^*$ \\
\hline
\end{tabular}
\end{center}
\subsection{$(\overline LL)(\overline RR)$ operators}
\begin{center}
\begin{tabular}{|c|c|}
\hline\textbf{LEFT WC} & \textbf{Matching} \\
\hline$\dfrac1{\Lambda^2}C_{\substack{\nu e\\prst}}^{V,LR}$ & $\dfrac1{\Lambda^2}\left[C_{\substack{le\\prst}}+\dfrac{{v_T}^2}{2\Lambda^2}\left(C_{\substack{l^2e^2H^2\\prst}}^{(1)}-C_{\substack{l^2e^2H^2\\prst}}^{(2)}\right)\right]-\dfrac{\overline{g_Z}^2}{{M_Z}^2}[Z_{\nu_L}]^{\text{eff}}_{pr}[Z_{e_R}]^{\text{eff}}_{st}$ \\[8pt]
\hline$\dfrac1{\Lambda^2}C_{\substack{ee\\prst}}^{V,LR}$ & $\dfrac1{\Lambda^2}\left[C_{\substack{le\\prst}}+\dfrac{{v_T}^2}{2\Lambda^2}\left(C_{\substack{l^2e^2H^2\\prst}}^{(1)}+C_{\substack{l^2e^2H^2\\prst}}^{(2)}\right)\right]$ \\[12pt]
& $-\dfrac{\overline{g_Z}^2}{{M_Z}^2}[Z_{e_L}]^{\text{eff}}_{pr}[Z_{e_R}]^{\text{eff}}_{st}-\dfrac1{2{m_h}^2}\,(Y_e)^{\text{eff}}_{pt}{(Y_e)^{\text{eff}}_{rs}}^*$ \\[8pt]
\hline$\dfrac1{\Lambda^2}C_{\substack{\nu u\\prst}}^{V,LR}$ & $\dfrac1{\Lambda^2}\left[C_{\substack{lu\\prst}}+\dfrac{{v_T}^2}{2\Lambda^2}\left(C_{\substack{l^2u^2H^2\\prst}}^{(1)}-C_{\substack{l^2u^2H^2\\prst}}^{(2)}\right)\right]-\dfrac{\overline{g_Z}^2}{{M_Z}^2}[Z_{\nu_L}]^{\text{eff}}_{pr}[Z_{u_R}]^{\text{eff}}_{st}$ \\[8pt]
\hline$\dfrac1{\Lambda^2}C_{\substack{eu\\prst}}^{V,LR}$ & $\dfrac1{\Lambda^2}\left[C_{\substack{lu\\prst}}+\dfrac{{v_T}^2}{2\Lambda^2}\left(C_{\substack{l^2u^2H^2\\prst}}^{(1)}+C_{\substack{l^2u^2H^2\\prst}}^{(2)}\right)\right]-\dfrac{\overline{g_Z}^2}{{M_Z}^2}[Z_{e_L}]^{\text{eff}}_{pr}[Z_{u_R}]^{\text{eff}}_{st}$ \\[8pt]
\hline$\dfrac1{\Lambda^2}C_{\substack{\nu d\\prst}}^{V,LR}$ & $\dfrac1{\Lambda^2}\left[C_{\substack{ld\\prst}}+\dfrac{{v_T}^2}{2\Lambda^2}\left(C_{\substack{l^2d^2H^2\\prst}}^{(1)}-C_{\substack{l^2d^2H^2\\prst}}^{(2)}\right)\right]-\dfrac{\overline{g_Z}^2}{{M_Z}^2}[Z_{\nu_L}]^{\text{eff}}_{pr}[Z_{d_R}]^{\text{eff}}_{st}$ \\
\hline
\end{tabular}
\end{center}
\begin{center}
\begin{tabular}{|c|c|}
\hline\textbf{LEFT WC} & \textbf{Matching} \\
\hline$\dfrac1{\Lambda^2}C_{\substack{uu\\prst}}^{V1,LR}$ & $\dfrac1{\Lambda^2}\left[C^{(1)}_{\substack{qu\\prst}}+\dfrac{{v_T}^2}{2\Lambda^2}\left(C_{\substack{q^2u^2H^2\\prst}}^{(1)}-C_{\substack{q^2u^2H^2\\prst}}^{(2)}\right)\right]$ \\[12pt]
& $-\dfrac{\overline{g_Z}^2}{{M_Z}^2}[Z_{u_L}]^{\text{eff}}_{pr}[Z_{u_R}]^{\text{eff}}_{st}-\dfrac1{6{m_h}^2}\,(Y_u)^{\text{eff}}_{pt}{(Y_u)^{\text{eff}}_{rs}}^*$ \\[8pt]
\hline$\dfrac1{\Lambda^2}C_{\substack{du\\prst}}^{V1,LR}$ & $\dfrac1{\Lambda^2}\left[C^{(1)}_{\substack{qu\\prst}}+\dfrac{{v_T}^2}{2\Lambda^2}\left(C_{\substack{q^2u^2H^2\\prst}}^{(1)}+C_{\substack{q^2u^2H^2\\prst}}^{(2)}\right)\right]-\dfrac{\overline{g_Z}^2}{{M_Z}^2}[Z_{d_L}]^{\text{eff}}_{pr}[Z_{u_R}]^{\text{eff}}_{st}$ \\[12pt]
\hline$\dfrac1{\Lambda^2}C_{\substack{uu\\prst}}^{V8,LR}$ & $\dfrac1{\Lambda^2}\left[C^{(8)}_{\substack{qu\\prst}}+\dfrac{{v_T}^2}{2\Lambda^2}\left(C_{\substack{q^2u^2H^2\\prst}}^{(3)}-C_{\substack{q^2u^2H^2\\prst}}^{(4)}\right)\right]-\dfrac1{{m_h}^2}\,(Y_u)^{\text{eff}}_{pt}{(Y_u)^{\text{eff}}_{rs}}^*$ \\[12pt]
\hline$\dfrac1{\Lambda^2}C_{\substack{du\\prst}}^{V8,LR}$ & $\dfrac1{\Lambda^2}\left[C^{(8)}_{\substack{qu\\prst}}+\dfrac{{v_T}^2}{2\Lambda^2}\left(C_{\substack{q^2u^2H^2\\prst}}^{(3)}+C_{\substack{q^2u^2H^2\\prst}}^{(4)}\right)\right]$ \\[12pt]
\hline$\dfrac1{\Lambda^2}C_{\substack{ud\\prst}}^{V1,LR}$ & $\dfrac1{\Lambda^2}\left[C^{(1)}_{\substack{qd\\prst}}+\dfrac{{v_T}^2}{2\Lambda^2}\left(C_{\substack{q^2d^2H^2\\prst}}^{(1)}-C_{\substack{q^2d^2H^2\\prst}}^{(2)}\right)\right]-\dfrac{\overline{g_Z}^2}{{M_Z}^2}[Z_{u_L}]^{\text{eff}}_{pr}[Z_{d_R}]^{\text{eff}}_{st}$ \\[12pt]
\hline$\dfrac1{\Lambda^2}C_{\substack{dd\\prst}}^{V1,LR}$ & $\dfrac1{\Lambda^2}\left[C^{(1)}_{\substack{qd\\prst}}+\dfrac{{v_T}^2}{2\Lambda^2}\left(C_{\substack{q^2d^2H^2\\prst}}^{(1)}+C_{\substack{q^2d^2H^2\\prst}}^{(2)}\right)\right]$ \\
\hline
\end{tabular}
\end{center}
\subsection{$(\overline LR)(\overline LR)$ operators}
\begin{center}
\begin{tabular}{|c|c|}
\hline\textbf{LEFT WC ($+$c.c.)} & \textbf{Matching ($+$c.c.)} \\
\hline$\dfrac1{\Lambda^2}C_{\substack{ee\\prst}}^{S,RR}$ & $\dfrac{{v_T}^2}{2\Lambda^4}\,C_{\substack{l^2e^2H^2\\prst}}^{(3)}+\dfrac1{2{m_h}^2}\,(Y_e)^{\text{eff}}_{pr}(Y_e)^{\text{eff}}_{st}$ \\[8pt]
\hline$\dfrac1{\Lambda^2}C_{\substack{eu\\prst}}^{S,RR}$ & $\dfrac1{\Lambda^2}\left[-C^{(1)}_{\substack{lequ\\prst}}+\dfrac{{v_T}^2}{2\Lambda^2}\left(-C_{\substack{lequH^2\\prst}}^{(1)}-C_{\substack{lequH^2\\prst}}^{(2)}\right)\right]+\dfrac1{{m_h}^2}\,(Y_e)^{\text{eff}}_{pr}(Y_u)^{\text{eff}}_{st}$ \\[12pt]
\hline$\dfrac1{\Lambda^2}C_{\substack{eu\\prst}}^{T,RR}$ & $\dfrac1{\Lambda^2}\left[-C^{(3)}_{\substack{lequ\\prst}}+\dfrac{{v_T}^2}{2\Lambda^2}\left(-C_{\substack{lequH^2\\prst}}^{(3)}-C_{\substack{lequH^2\\prst}}^{(4)}\right)\right]$ \\[12pt]
\hline$\dfrac1{\Lambda^2}C_{\substack{ed\\prst}}^{S,RR}$ & $\dfrac{{v_T}^2}{2\Lambda^4}C_{\substack{leqdH^2\\prst}}^{(3)}+\dfrac1{{m_h}^2}\,(Y_e)^{\text{eff}}_{pr}(Y_d)^{\text{eff}}_{st}$ \\[8pt]
\hline$\dfrac1{\Lambda^2}C_{\substack{ed\\prst}}^{T,RR}$ & $\dfrac{{v_T}^2}{2\Lambda^4}C_{\substack{leqdH^2\\prst}}^{(4)}$ \\[8pt]
\hline$\dfrac1{\Lambda^2}C_{\substack{\nu edu\\prst}}^{S,RR}$ & $\dfrac1{\Lambda^2}\left[C^{(1)}_{\substack{lequ\\prst}}+\dfrac{{v_T}^2}{2\Lambda^2}\left(C_{\substack{lequH^2\\prst}}^{(1)}-C_{\substack{lequH^2\\prst}}^{(2)}\right)\right]$ \\[12pt]
\hline$\dfrac1{\Lambda^2}C_{\substack{\nu edu\\prst}}^{T,RR}$ & $\dfrac1{\Lambda^2}\left[C^{(3)}_{\substack{lequ\\prst}}+\dfrac{{v_T}^2}{2\Lambda^2}\left(C_{\substack{lequH^2\\prst}}^{(3)}-C_{\substack{lequH^2\\prst}}^{(4)}\right)\right]$ \\
\hline
\end{tabular}
\end{center}
\subsection{$(\overline LR)(\overline RL)+\,$ h.c. operators}
\begin{center}
\begin{tabular}{|c|c|}
\hline\textbf{LEFT WC ($+\,$c.c.)} & \textbf{Matching ($+\,$c.c.)} \\
\hline$\dfrac1{\Lambda^2}C_{\substack{eu\\prst}}^{S,RL}$ & $\dfrac{{v_T}^2}{2\Lambda^4}\,C_{\substack{lequH^2\\prst}}^{(5)}+\dfrac1{{m_h}^2}\,(Y_e)^{\text{eff}}_{pr}{(Y_u)^{\text{eff}}_{ts}}^*$ \\[8pt]
\hline$\dfrac1{\Lambda^2}C_{\substack{ed\\prst}}^{S,RL}$ & $\dfrac1{\Lambda^2}\left[C_{\substack{ledq\\prst}}+\dfrac{{v_T}^2}{2\Lambda^2}\,\left(C_{\substack{leqdH^2\\prst}}^{(1)}+C_{\substack{leqdH^2\\prst}}^{(2)}\right)\right]+\dfrac1{{m_h}^2}\,(Y_e)^{\text{eff}}_{pr}{(Y_d)^{\text{eff}}_{ts}}^*$ \\[12pt]
\hline$\dfrac1{\Lambda^2}C_{\substack{\nu edu\\prst}}^{S,RL}\,+$ h.c. & $\dfrac1{\Lambda^2}\left[C_{\substack{ledq\\prst}}+\dfrac{{v_T}^2}{2\Lambda^2}\left(C_{\substack{leqdH^2\\prst}}^{(1)}-C_{\substack{leqdH^2\\prst}}^{(2)}\right)\right]$ \\[12pt]
\hline
\end{tabular}
\end{center}
\subsection{$\Delta L=4+\,$ h.c. operator}
\begin{center}
\begin{tabular}{|c|c|}
\hline\textbf{LEFT WC ($+\,$c.c.)} & \textbf{Matching ($+\,$c.c.)} \\
\hline$\dfrac1{\Lambda^2}C_{\substack{\nu\nu\\prst}}^{S,LL}$ & $0$ \\
\hline
\end{tabular}
\end{center}
\subsection{$\Delta L=2+\,$ h.c. operators}
\begin{center}
\begin{tabular}{|c|c|}
\hline\textbf{LEFT WC ($+\,$c.c.)} & \textbf{Matching ($+\,$c.c.)} \\
\hline$\dfrac1{\Lambda^2}C_{\substack{\nu e\\prst}}^{S,LL}$ & $\dfrac{v_T}{2\sqrt2\,\Lambda^3}\left[\left(C_{\substack{l^3eH\\prst}}+C_{\substack{l^3eH\\rpst}}\right)+\dfrac12\left(C_{\substack{l^3eH\\tpsr}}+C_{\substack{l^3eH\\trsp}}\right)\right]$ \\[8pt]
\hline$\dfrac1{\Lambda^2}C_{\substack{\nu e\\prst}}^{T,LL}$ & $\dfrac{v_T}{16\sqrt2\,\Lambda^3}\left(C_{\substack{l^3eH\\tpsr}}-C_{\substack{l^3eH\\trsp}}\right)$ \\[8pt]
\hline$\dfrac1{\Lambda^2}C_{\substack{\nu e\\prst}}^{S,LR}$ & $\dfrac{\overline g^2}{2{M_W}^2}\left([W_l^{\slashed L}]^{\text{eff}}_{pt}{[W_l]^{\text{eff}}_{rs}}^*+[W_l^{\slashed L}]^{\text{eff}}_{rt}{[W_l]^{\text{eff}}_{ps}}^*\right)$ \\[8pt]
\hline$\dfrac1{\Lambda^2}C_{\substack{\nu u\\prst}}^{S,LL}$ & $0$ \\[8pt]
\hline$\dfrac1{\Lambda^2}C_{\substack{\nu u\\prst}}^{T,LL}$ & $0$ \\[8pt]
\hline$\dfrac1{\Lambda^2}C_{\substack{\nu u\\prst}}^{S,LR}$ & $\dfrac{v_T}{2\sqrt2\,\Lambda^3}\left(C_{\substack{l^2quH\\prst}}+C_{\substack{l^2quH\\rpst}}\right)$ \\[8pt]
\hline$\dfrac1{\Lambda^2}C_{\substack{\nu d\\prst}}^{S,LL}$ & $\dfrac{v_T}{2\sqrt2\,\Lambda^3}\left(C^{(1)}_{\substack{l^2dqH\\prst}}+C^{(1)}_{\substack{l^2dqH\\rpst}}\right)$ \\
\hline
\end{tabular}
\end{center}
\subsection{$\Delta B=\Delta L=1+\,$ h.c. operators}
\begin{center}
\begin{tabular}{|c|c|}
\hline\textbf{LEFT WC ($+\,$c.c.)} & \textbf{Matching ($+\,$c.c.)} \\
\hline$\dfrac1{\Lambda^2}C_{\substack{udd\\prst}}^{S,LL}$ & $\dfrac1{\Lambda^2}\left[\begin{aligned}
& \left(C_{\substack{qqq\\rpst}}+C_{\substack{qqq\\srpt}}-C_{\substack{qqq\\rspt}}\right) \\
& +\dfrac{{v_T}^2}{2\Lambda^2}\left(\begin{aligned}
& C^{(1)}_{\substack{lq^3H^2\\rpst}}+C^{(1)}_{\substack{lq^3H^2\\srpt}}-C^{(1)}_{\substack{lq^3H^2\\rspt}}+C^{(2)}_{\substack{lq^3H^2\\rpst}}+C^{(2)}_{\substack{lq^3H^2\\srpt}} \\
& -C^{(2)}_{\substack{lq^3H^2\\rspt}}-C^{(3)}_{\substack{lq^3H^2\\rpst}}-C^{(3)}_{\substack{lq^3H^2\\srpt}}+C^{(3)}_{\substack{lq^3H^2\\rspt}}
\end{aligned}\right)
\end{aligned}\right]$ \\[38pt]
\hline$\dfrac1{\Lambda^2}C_{\substack{duu\\prst}}^{S,LL}$ & $\dfrac1{\Lambda^2}\left[\begin{aligned}
& \left(C_{\substack{qqq\\rpst}}+C_{\substack{qqq\\srpt}}-C_{\substack{qqq\\rspt}}\right) \\
& +\dfrac{{v_T}^2}{2\Lambda^2}\left(\begin{aligned}
& C^{(1)}_{\substack{lq^3H^2\\rpst}}+C^{(1)}_{\substack{lq^3H^2\\srpt}}-C^{(1)}_{\substack{lq^3H^2\\rspt}}-C^{(2)}_{\substack{lq^3H^2\\rpst}}-C^{(2)}_{\substack{lq^3H^2\\srpt}} \\
& +C^{(2)}_{\substack{lq^3H^2\\rspt}}+C^{(3)}_{\substack{lq^3H^2\\rpst}}+C^{(3)}_{\substack{lq^3H^2\\srpt}}-C^{(3)}_{\substack{lq^3H^2\\rspt}}
\end{aligned}\right)
\end{aligned}\right]$ \\[38pt]
\hline$\dfrac1{\Lambda^2}C_{\substack{uud\\prst}}^{S,LR}$ & $\dfrac{{v_T}^2}{2\Lambda^4}C_{\substack{eq^2dH^2\\tspr}}$ \\[8pt]
\hline$\dfrac1{\Lambda^2}C_{\substack{duu\\prst}}^{S,LR}$ & $-\dfrac1{\Lambda^2}\left[\left(C_{\substack{qqu\\prst}}+C_{\substack{qqu\\rpst}}\right)+\dfrac{{v_T}^2}{2\Lambda^2}C_{\substack{eq^2uH^2\\rpst}}\right]$ \\[8pt]
\hline$\dfrac1{\Lambda^2}C_{\substack{uud\\prst}}^{S,RL}$ & $\dfrac{{v_T}^2}{2\Lambda^4}C_{\substack{lqu^2H^2\\tspr}}$ \\
\hline
\end{tabular}
\end{center}
\subsection{$\Delta B=-\Delta L=1+\,$ h.c. operators}
\begin{center}
\begin{tabular}{|c|c|c|}
\hline\textbf{LEFT WC ($+\,$c.c.)} & \textbf{Matching ($+\,$c.c.)} \\
\hline$\dfrac1{\Lambda^2}C_{\substack{ddd\\prst}}^{S,LL}$ & $0$ \\[8pt]
\hline$\dfrac1{\Lambda^2}C_{\substack{udd\\prst}}^{S,LR}$ & $-\dfrac{v_T}{\sqrt2\,\Lambda^3}C_{\substack{q^2ldH\\prst}}$ \\[8pt]
\hline$\dfrac1{\Lambda^2}C_{\substack{ddu\\prst}}^{S,LR}$ & $0$ \\[8pt]
\hline$\dfrac1{\Lambda^2}C_{\substack{ddd\\prst}}^{S,LR}$ & $-\dfrac{v_T}{2\sqrt2\,\Lambda^3}\left(C_{\substack{q^2ldH\\prst}}-C_{\substack{q^2ldH\\rpst}}\right)$ \\[8pt]
\hline$\dfrac1{\Lambda^2}C_{\substack{ddd\\prst}}^{S,RL}$ & $-\dfrac{v_T}{\sqrt2\,\Lambda^3}C_{\substack{eqd^2H\\prst}}$ \\[8pt]
\hline$\dfrac1{\Lambda^2}C_{\substack{udd\\prst}}^{S,RR}$ & $\dfrac{v_T}{\sqrt2\,\Lambda^3}C_{\substack{ud^2lH\\prst}}$ \\[8pt]
\hline$\dfrac1{\Lambda^2}C_{\substack{ddd\\prst}}^{S,RR}$ & $\dfrac{v_T}{\sqrt2\,\Lambda^3}C_{\substack{d^3lH\\prst}}$ \\[8pt]
\hline
\end{tabular}
\end{center}

\end{document}